\documentclass{eas}
\usepackage{graphicx}
\TitreGlobal{What can the highest angular resolution bring to stellar astrophysics?}
\begin{document}

\title{Stellar granulation and interferometry} 
\newcommand{\folder}{.}

%
\author{A. Chiavassa$^1$} \author{L. Bigot} 
\address{Laboratoire Lagrange, UMR 7293, CNRS, Observatoire de la C\^ote d'Azur, Universit\'e de Nice Sophia-Antipolis, Nice, France}

%
%
\begin{abstract}
Stars are not smooth. Their photosphere is covered by a granulation
pattern associated with the heat transport by convection. The
convection-related surface structures have different size, depth, and
temporal variations with respect to the stellar type. The related
activity (in addition to other phenomena such as magnetic spots,
rotation, dust, etc.) potentially causes bias in stellar parameters
determination, radial velocity, chemical abundances determinations,
and exoplanet transit detections.

The role of long-baseline interferometric observations in this
astrophysical context is crucial to characterize the stellar surface
dynamics and correct the potential biases. In this Chapter, we present
how the granulation pattern is expected for different kind of stellar
types ranging from main sequence to extremely evolved stars of
different masses and how interferometric techniques help to study
their photospheric dynamics.
\end{abstract}
\maketitle
\section{Introduction}

Stellar granulation was observed for the first time on the Sun by
Herschel (\cite{1801RSPT...91..265H}), but it was Dawes
(\cite{1864MNRAS..24..161D}) who coined the term granules. Eventually,
today modern telescopes provide direct observations (e.g., Carlsson
{\em et al.\/}, \cite{2004ApJ...610L.137C}).  The granulation pattern
is associated with heat transport by convection, on horizontal scales
of the order of a thousand kilometers (Nordlund {\em et
al.\/}, \cite{2009LRSP....6....2N}). On the other hand, convection is
driven primarily by radiative cooling from a thin thermal boundary
layer, the layer from which most photons can escape to space. The most
prominent intensity variations on the solar surface, aside from
sunspots and faculae, are granules i.e. the bright (hot) areas
surrounded by dark (cooler) lanes that tile the stellar surface. The
horizontal scale on which radiative cooling drives the convective
motions is linked with the granulation diameter. The bright granules
are the locations of upflowing hot plasma, while the dark
intergranular lanes are the locations of downflowing cool plasma
(Nordlund {\em et al.\/}, \cite{1990A&A...228..155N}).

Stellar granulation manifests either on resolved observables (e.g.,
Sun images) or unresolved photospheric spectral line in terms of
widths, shapes, and strengths. In particular, the best observational
evidence comes from unresolved spectral lines because they combine, in
their shapes, important properties such as velocity amplitudes
(heavily affecting the line width) and velocity-intensity correlations
(Nordlund {\em et al.\/}, \cite{2009LRSP....6....2N}). Similarly,
correlation of velocity and temperature cause characteristic
asymmetries of spectral lines as well as net blueshifts
(Dravins, \cite{1987A&A...172..211D};
Gray, \cite{2005oasp.book.....G}).

Stellar granulation potentially cause bias in stellar parameters,
radial velocity, chemical abundances determinations, and exoplanet
transit detections.  For this purpose, large efforts have been made in
recent decades to use theoretical modelling of stellar atmospheres to
solve multidimensional radiative hydrodynamic equations in which
convection emerges naturally. These simulations take surface
inhomogeneities into account (e.g., granulation pattern) and velocity
fields and are used to predict reliable observables. They cover a
substantial portion of the Hertzsprung-Russell diagram (Magic {\em et
al.\/}, \cite{2013A&A...557A..26M}; Ludwig {\em et
al.\/}, \cite{2009MmSAI..80..711L}), including the evolutionary phases
from the main sequence over the turnoff up to the red-giant branch for
low-mass stars. These simulations of stellar atmospheres are essential
for much of contemporary astronomy.

\section[3D RHD simulations]{Stellar atmosphere, a pathway to multi-dimensional simulations}\label{Sectmodel}

The primary source of information for stellar objects is the light
they emit, which carries information about the physical conditions at
its origin. However, in order to interpret the information correctly,
one first needs either theoretical or semi-empirical models of the
atmospheric layers at the surface of stars from where the stellar
radiation escapes. Many of the observable phenomena occurring on the
surface of the stars are intimately linked to convection. Moreover,
the atmospheric temperature stratification in the optically thin region,
where the emerging flux form, is also affected by the interaction
between radiative and convective energy transport. To account for all
these aspects, it is important to use realistic 3D radiative
hydrodynamical (RHD) simulations of stellar convection (Nordlund {\em
et al.\/} \cite{1998ApJ...499..914S}; Freytag {\em et
al.\/} \cite{2012JCoPh.231..919F}; Ludwig {\em et
al.\/} \cite{2004ApJ...610L.137C}; V{\"o}gler {\em et
al.\/}, \cite{2004A&A...421..741V}).

RHD simulations numerically solves the time-dependent equations for
conservation of mass, momentum, and energy coupled to a realistic
treatment of the radiative transfer. The simulation domains are of two
kinds: (i) \emph{box-in-a-star} simulations (Fig.~\ref{scheme}, right
panel; computational time ranging from few days to few weeks depending
on the stellar type) cover only a small section of the surface layers
of the deep convection zone (typically ten pressure scale heights
vertically), and the numerical box includes about $\sim10$ convective
cells, which are large enough so that the cells are not constrained by
the horizontal (cyclic) boundaries. (ii) \emph{star-in-a-box}
simulations (Fig.~\ref{scheme}, left panel; computational time of few
months) cover the whole convective envelope of the star and have been
used to model evolved cool stars like Red Supergiant stars, RSGs
(Freytag {\em et al.\/}, \cite{2012JCoPh.231..919F}; Chiavassa {\em et
al.\/}, \cite{2011A&A...535A..22C}) and Asymptotic Giant Branch stars,
AGBs (Freytag {\em et al.\/}, \cite{2008A&A...483..571F}) so far. Once
the surface gravity is lower than $\log g \sim 1$, the box-in-a-star
simulations become inadequate because of the influence of sphericity
becomes important; the star-in-a-box global simulations are then
needed, but those are highly computer-time demanding and difficult to
run, which is the reason why there are only very few simulations
available so far.

RHD simulations employ realistic input physics: (i) updated and
reliable equation of state, which characterize the thermodynamic state
of the matter; (ii) the radiative transfer includes the most recent
continuous absorption coefficients as well as atomic and molecular
line opacities. The abundances employed in the computation of the
simulation are the latest chemical composition by Ashland {\em et
al.\/} (\cite{2009ARA&A..47..481A}).

In particular, the radiation transport used in RHD simulations is a
crucial ingredient either for the resulting atmospheric stratification
and the time needed for the computation. It cannot be computed
monochromatically for all the wavelength across the spectrum because
of the very large computational time needed. Therefore The frequency
dependance of the radiation field can be calculated with two
approaches:
\begin{itemize}
\item  the gray approximation, which completely ignores the frequency
       dependence, is justified only in the stellar interior and it is
       inaccurate in the optically thin layers. This method is often
       used for \emph{star-in-a-box} simulations.
\item the more elaborate scheme accounting for non-gray effects, which
      is based on the idea of \emph{opacity binning}
      (Nordlund, \cite{1982A&A...107....1N}; Nordlund {\em et
      al.\/},\cite{1990A&A...228..155N}). The basic approximation is
      the so called multi-group scheme (Ludwig {\em et
      al.\/}, \cite{1994A&A...284..105L}; V{\"o}gler {\em et
      al.\/} \cite{2004A&A...421..741V}). In this scheme, the
      frequencies that reach monochromatic optical depth unity within
      a certain depth range of the model atmosphere will be put into
      one frequency group. Actual RHD simulations use up to 5 and 12
      bins (\emph{star-in-a-box} and \emph{box-in-a-star} simulations,
      respectively) to reproduce the whole frequency dependence across
      the full spectrum
\end{itemize}

\section[Radiative transfer]{Detailed radiative transfer to extract observables}

\begin{figure}
\centering
  \begin{tabular}{cc}
   \includegraphics[width=0.47\hsize]{\folder/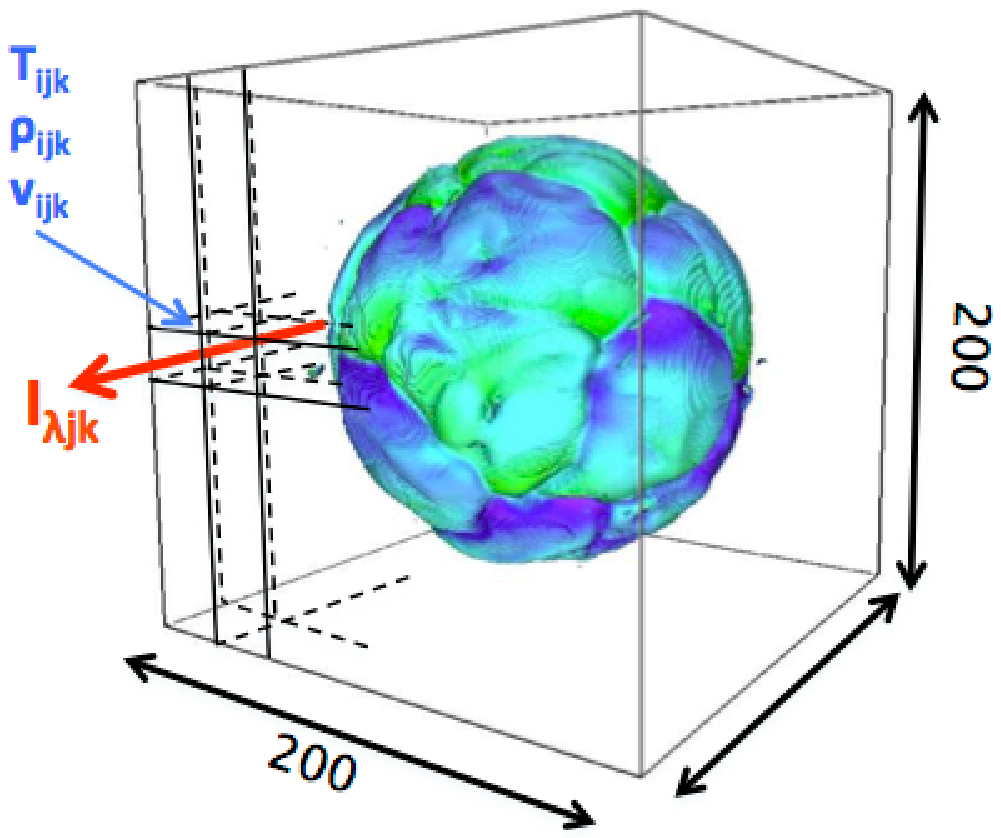}
   \includegraphics[width=0.45\hsize]{\folder/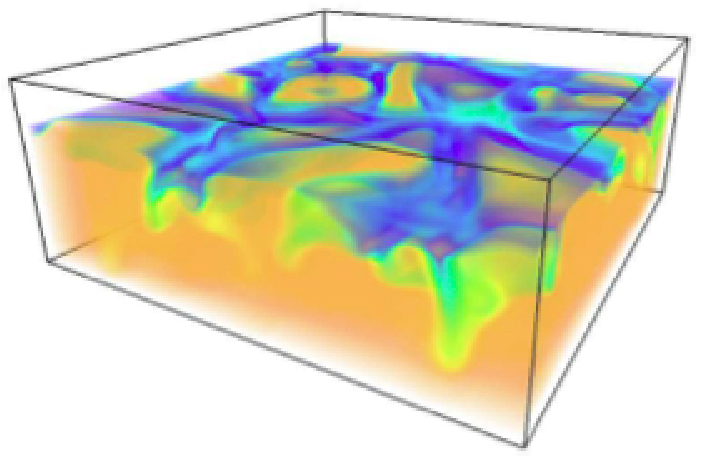}\\
  \end{tabular}
      \caption{\emph{Left panel:} Scheme of the interaction between {{\sc Optim3D}} and RHD simulations. The RHD simulation (in the \emph{star-in-a-box} configuration, green to blue star in the middle of the numerical cube) thermodynamical variables (temperature, T$_{ijk}$; density, $\rho_{ink}$; and velocity, v$_{ink}$) are used in {{\sc Optim3D}} to compute the monochromatic emerging intensity ($I_{\lambda jk}$) along different casting rays. \emph{Right panel:} RHD simulation in the \emph{box-in-a-star} configuration.}
   \label{scheme}
   \end{figure}

The extraction of reliable observables require detailed monochromatic
calculations. To overcome the small number of bins in RHD simulations,
Chiavassa {\em et al.\/} (\cite{2009A&A...506.1351C}) developed a 3D
pure-LTE radiative transfer code called {{\sc Optim3D}} that make
computation between 1000 to 200000 \AA , with any spectral
resolution. The code takes into account the Doppler shifts caused by
convective motions. The radiative transfer equation is solved
monochromatically using extinction coefficients pre-tabulated as a
function of temperature, density, and wavelength (Fig.~\ref{scheme},
left panel). The lookup tables are computed for any kind of chemical
compositions using the same extensive atomic and molecular opacity
data as the latest generation of MARCS models (Gustafsson {\em et
al.\/}, \cite{2008A&A...486..951G}). {{\sc Optim3D}} assumes a zero
micro-turbulence (i.e., microscale, smaller than the free photon path,
non-thermal component of the gas velocity in the region of spectral
line formation.) because the velocity fields inherent in RHD
simulations are expected to self-consistently and adequately account
for non-thermal Doppler broadening of spectral lines.

The synergy between RHD simulations and {{\sc Optim3D}} allows to
tackle several astrophysical problems and is based on the systematic
production of observables like: high and low resolution spectra to
study the chemical abundances, velocity field, and pulsations of stars
as well as their stellar parameters; interferometric observables to
study the stellar dynamics, stellar parameters, planet transit; and
images to study the impact of convention on astrometric measurements.

\section[Where is the signal?]{Higher spatial frequencies for the stellar granulation signal}

\begin{figure}
\centering
  \begin{tabular}{cc}
   \includegraphics[width=0.47\hsize]{\folder/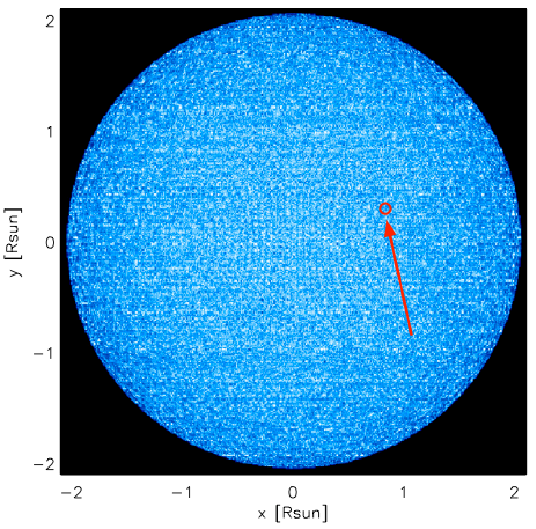}
   \includegraphics[width=0.45\hsize]{\folder/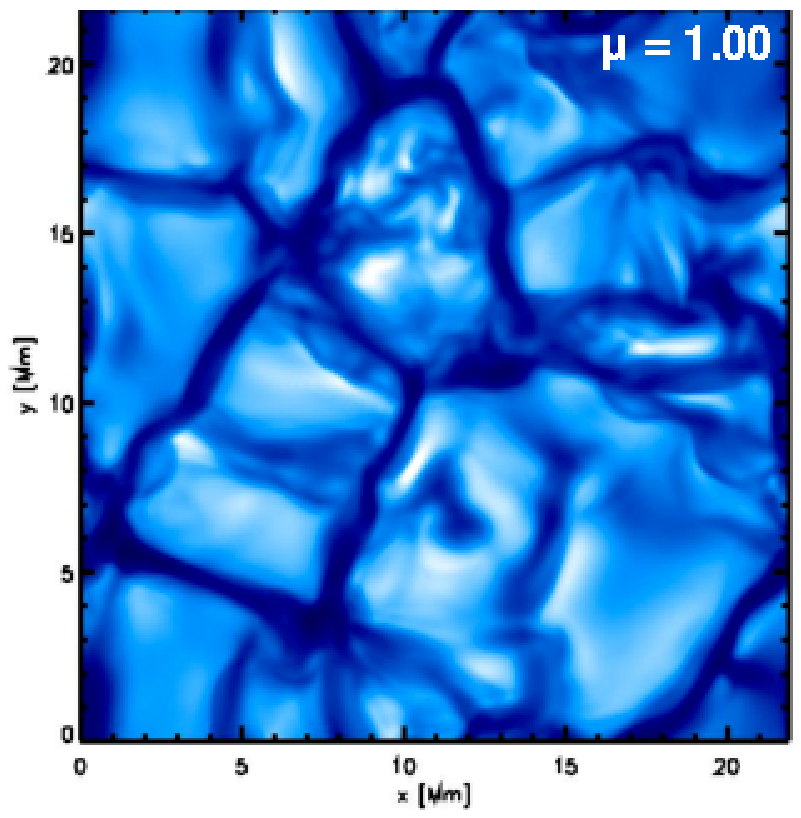}\\
   \includegraphics[width=0.47\hsize]{\folder/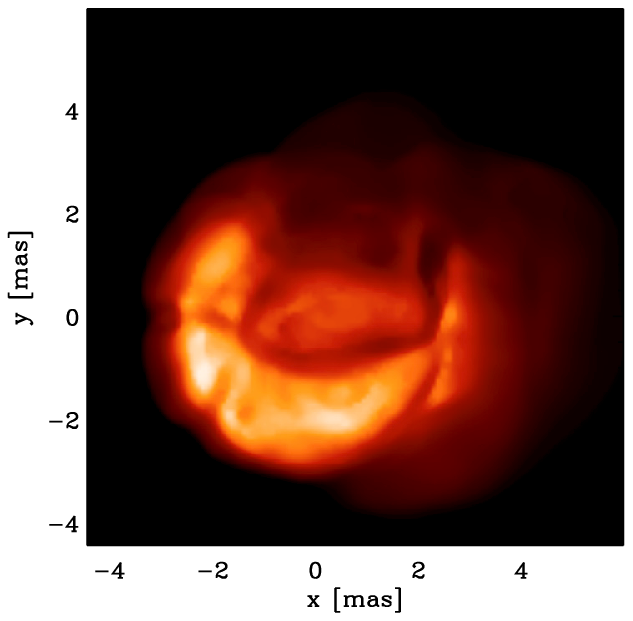}
   \includegraphics[width=0.51\hsize]{\folder/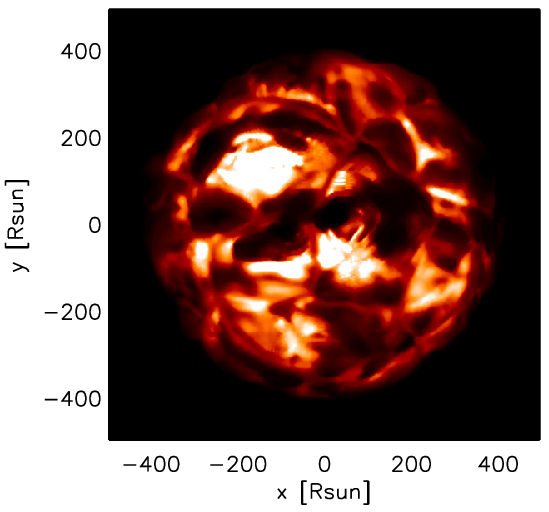}
\end{tabular}
      \caption{Synthetic images of different stars computed with {{\sc
               Optim3D}} at different wavelength and from RHD
               simulations. \emph{Top left panel:} Procyon with
               $T_{\mathrm{eff}}=6591 \pm 43 $ K, $\log g=4.01 \pm
               0.03$, radius $R = 2.023 \pm 0.026 R_{\odot}$, and mass
               (Chiavassa {\em et
               al.\/}, \cite{2012A&A...540A...5C}). The red circle and
               arrow indicate a particular zone used for the
               enlargement; \emph{Top right panel:} enlargement of the
               surface of Procyon. \emph{Bottom left panel:} AGB star
               with $T_{\mathrm{eff}}=2542 \pm 220 $ K, $\log
               g=-0.83 \pm 0.10$, $R=443.5 \pm 58.3 R_{\odot}$, and $M
               = 1 M_{\odot}$ (Freytag {\em et
               al.\/}, \cite{2008A&A...483..571F}; Chiavassa {\em et
               al.\/}, \cite{2010A&A...511A..51C}). \emph{Bottom right
               panel:} RSG star with $T_{\mathrm{eff}}=3710 \pm 20 $
               K, $\log g=0.047 \pm 0.001$, $R=376.7 \pm 0.5
               R_{\odot}$, and $M = 6 M_{\odot}$ (Chiavassa {\em et
               al.\/}, \cite{2011A&A...535A..22C}).}
  \label{granulation} \end{figure}

The size of the convective cells is linked to the pressure scale
height at optical-depth unity (Freytag {\em et
al.\/}, \cite{2001ASPC..223..785F}). The pressure scale height is
defined as

\begin{eqnarray}
\mathcal{H}_{\mathrm{p}}= \frac{k_B T_{\mathrm{eff}}}{mg},
\end{eqnarray}

where $g$ is the surface gravity, $k_B$ is the Boltzmann constant and
$m$ is the mean molecular mass ($m=1.31\times m_{\rm{H}} = 1.31 \times
1.67 \times10^{-24}$~grams, for temperatures lower than 10000~K). In
the above expression, $\mathcal{H}_{\mathrm{p}}$ has the dimension of
length. Stars with low surface gravity have more diluted atmosphere
and lower surface temperature while more compact objects are
hotter. Fig.~\ref{granulation} show the synthetic images of stars with
different stellar parameters and thus different granulation pattern.

Samadi {\em et al.\/} (\cite{2013A&A...559A..40S}), Svensson $\&$
Ludwig (\cite{2005ESASP.560..979S}) and Chiavassa {\em et al.\/}
(\cite{2011A&A...528A.120C}) reported a direct correlation between the
size of the convective cells and the stellar parameters: larger
granules have larger effects on the emerging stellar flux. Following
this principle, the role of long-baseline interferometric observations
is to investigate the dynamics of granulation as a function of stellar
parameters: thanks to the higher angular resolution, interferometry is
the ideal tool for exploring stellar convection in term of cell's
size, intensity contrast and temporal variations.

\subsection{Evolved stars}

\begin{figure}
   \centering
    \begin{tabular}{c}
        \includegraphics[width=0.97\hsize]{\folder/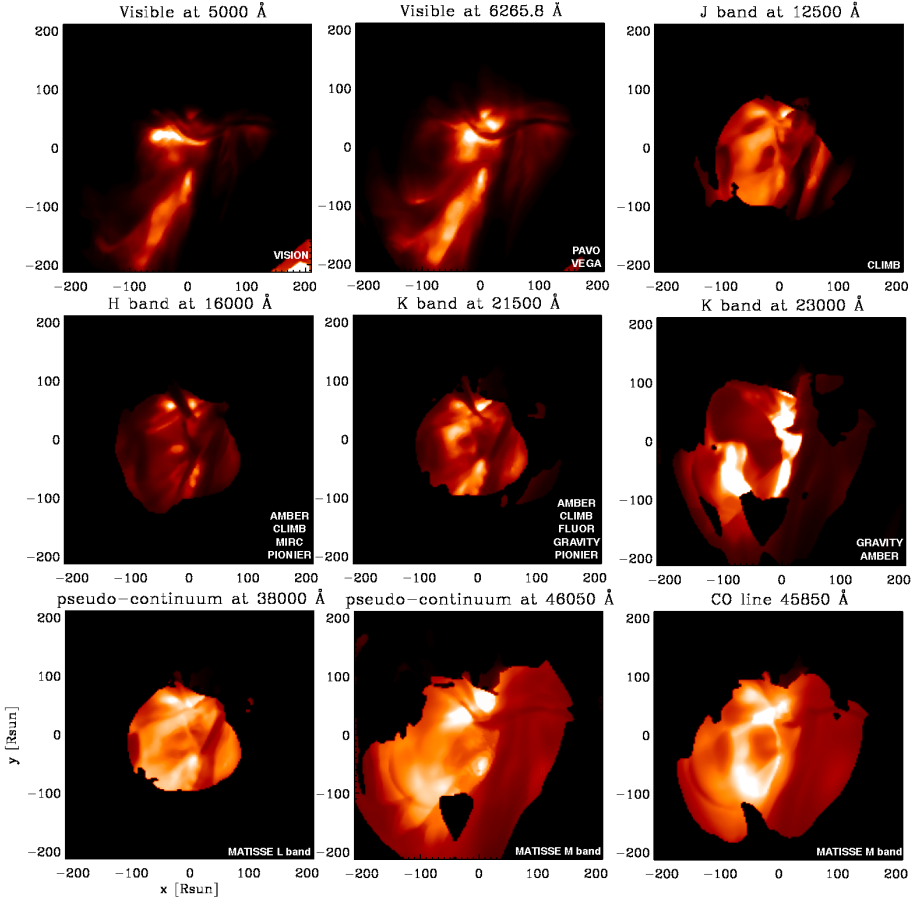}
   \end{tabular}
      \caption{Appearance of the stellar surface of an AGB simulation (image from Chiavassa and Freytag, \cite{2014arXiv1410.3868C}) with respect to the wavelength. The present and future interferometric instrument probing these wavelengths are indicated in lower right corner of every image.}
       \label{variability}
     \end{figure}

\begin{figure}
   \centering
    \begin{tabular}{ccc}
      \includegraphics[width=0.7\hsize]{\folder/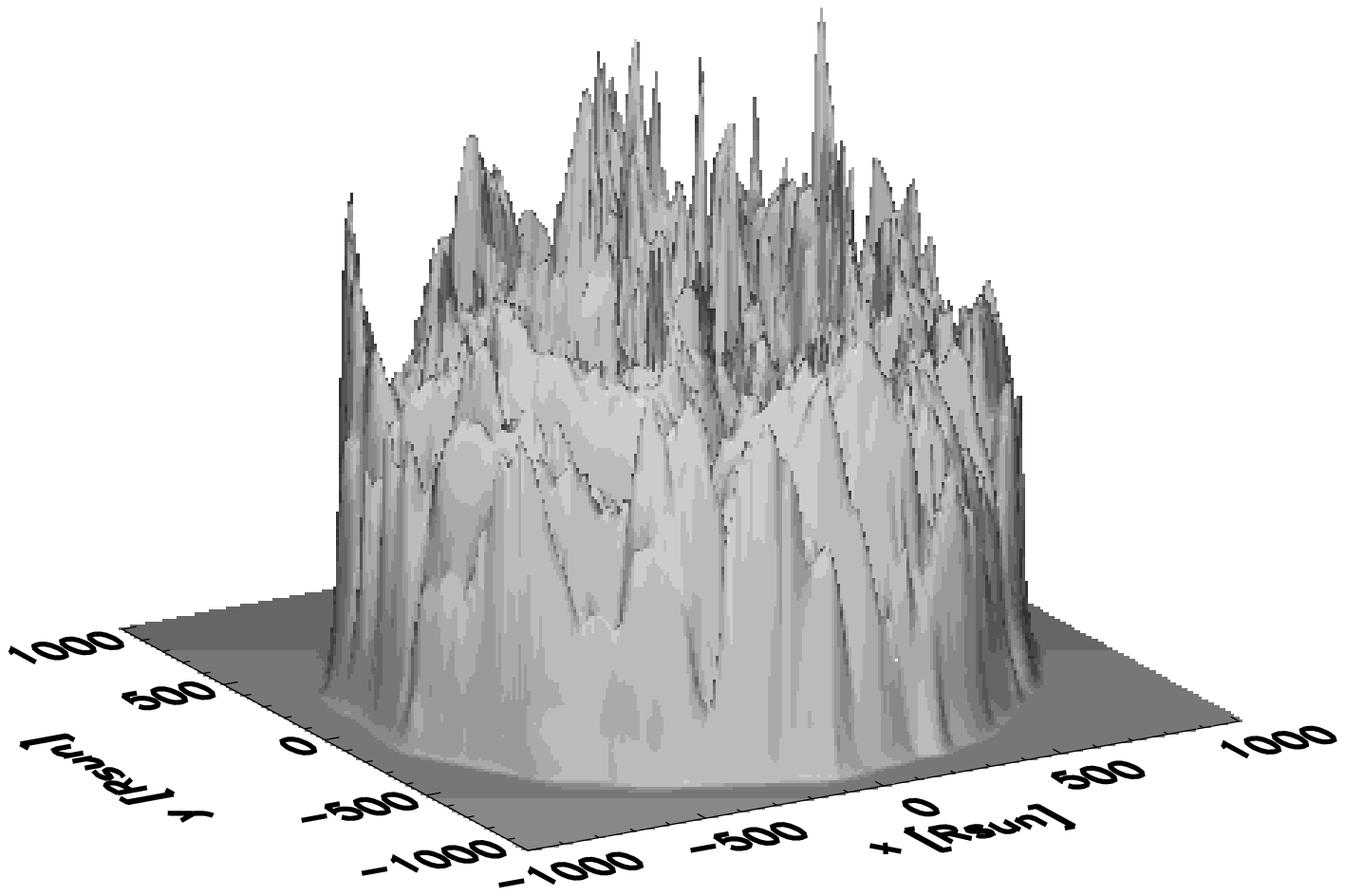} \\
      \includegraphics[width=0.5\hsize]{\folder/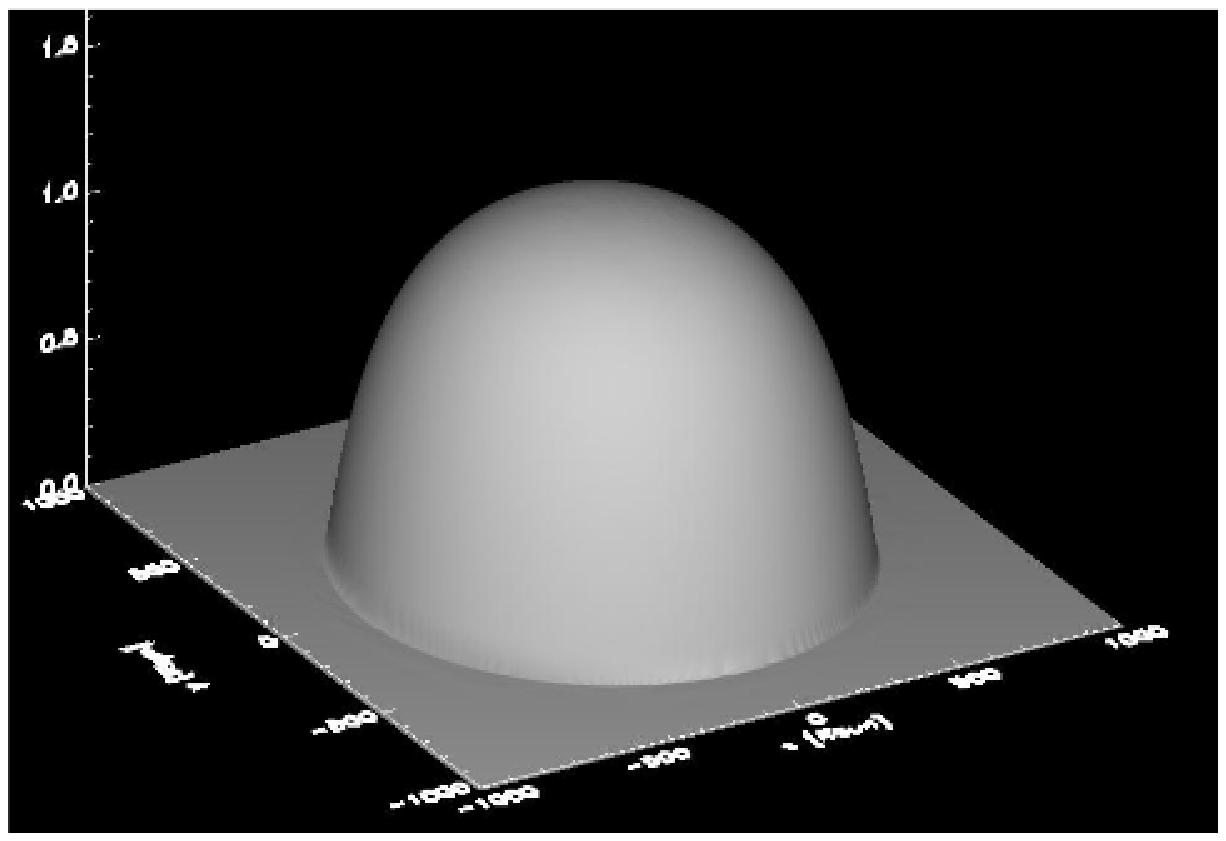}
      \includegraphics[width=0.48\hsize]{\folder/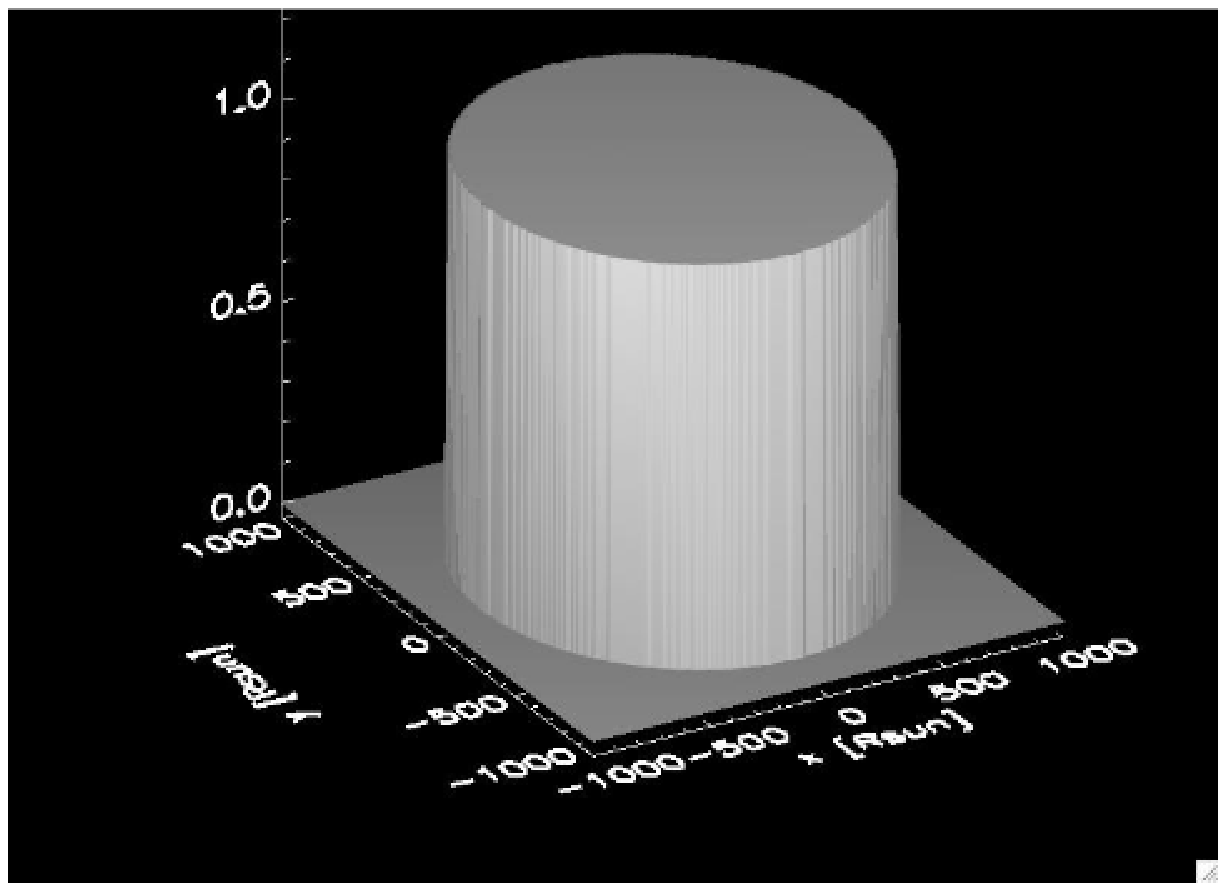}
 \end{tabular}
      \caption{\emph{Top panel: }three-dimensional image of the emerging intensity from a RHD simulation of RSG star in the H band (image from Chiavassa {\em et al.\/}, \cite{2009A&A...506.1351C}). \emph{Bottom left panel:} from a limb-darkened disk with the stellar parameter of the RHD simulation. \emph{Bottom right panel:} from a uniform disk.}
       \label{intensity_radial}
     \end{figure}

Evolved stars, such as RSGs and AGBs, are prime targets for
interferometry, because of their large diameter, proximity, and high
infrared luminosity. These stars have effective temperatures lower
than $\sim$ 4000 K and surface gravity lower than $\log g=1.0$. Their
atmospheres, resulting from the RHD simulations, is undoubtedly
irregular, permeated with structures and dynamics, but depend on the
wavelength probed (Fig.~\ref{granulation} and
Fig.~\ref{variability}). Moreover, the surface inhomogeneities and
their temporal evolution induce strong fluctuations on the intensity
profiles. Fig.~\ref{intensity_radial} displays the comparison between
a three-dimensional image representation of the intensity in a
snapshot of a RSG simulation in the H band with an ad-hoc
limb-darkened image representation for a RSG star and an uniform
disk. The difference is striking and the resulting RHD surface pattern
full of numerous intensity spikes. These, though related to the
underlying granulation pattern (characterized by very large convective
cells), are also connected to dynamical effects. The emerging
intensity depends on (i) the opacity run through the atmosphere (and
for $T_{\rm eff}<$4000 K, molecules produce strong absorption in
particular in the visible) and on (ii) the shocks and waves that
dominate at optical depths smaller than$\sim$1.

To search for the effect of the granulation pattern on interferometric
visibility curves and phases, we use the intensity maps computed with
{{\sc Optim3D}} (e.g., Fig.~\ref{granulation}) and calculated a
discrete Fourier transform ($FT$). We introduced also a theoretical spatial frequency
scale expressed in units of inverse solar radii (R$_\odot^{-1}$). The
conversion between spatial frequencies expressed in the latter scale
and in the more usual scale of $1/''$ is given by:

\begin{equation}\label{eqvis1}
\nu~\left[\frac{1}{''}\right]=\nu~\left[\frac{1}{{\rm R}_\odot^{-1}}\right]\cdot d~[{\rm pc}]\cdot214.9
\end{equation}
where $\nu$ is the spatial frequency, 
214.9  is the astronomical unit expressed in solar radii, and
$d$ is the distance of the observed star. Also useful is the following
relation 
\begin{equation}\label{eqvis2}
\nu=\frac{B}{\lambda\cdot0.206265}
\end{equation}
where $\nu$ is the spatial frequency in arcsec$^{-1}$ at the observed
wavelength $\lambda$ in $\mu{\rm m}$ for the baseline $B$ of an
interferometer in meters. 

\begin{figure}
   \centering
        \begin{tabular}{cc}
     \includegraphics[width=0.6\hsize]{\folder/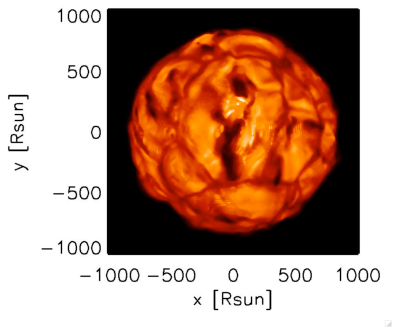}  \\
 \includegraphics[width=0.5\hsize]{\folder/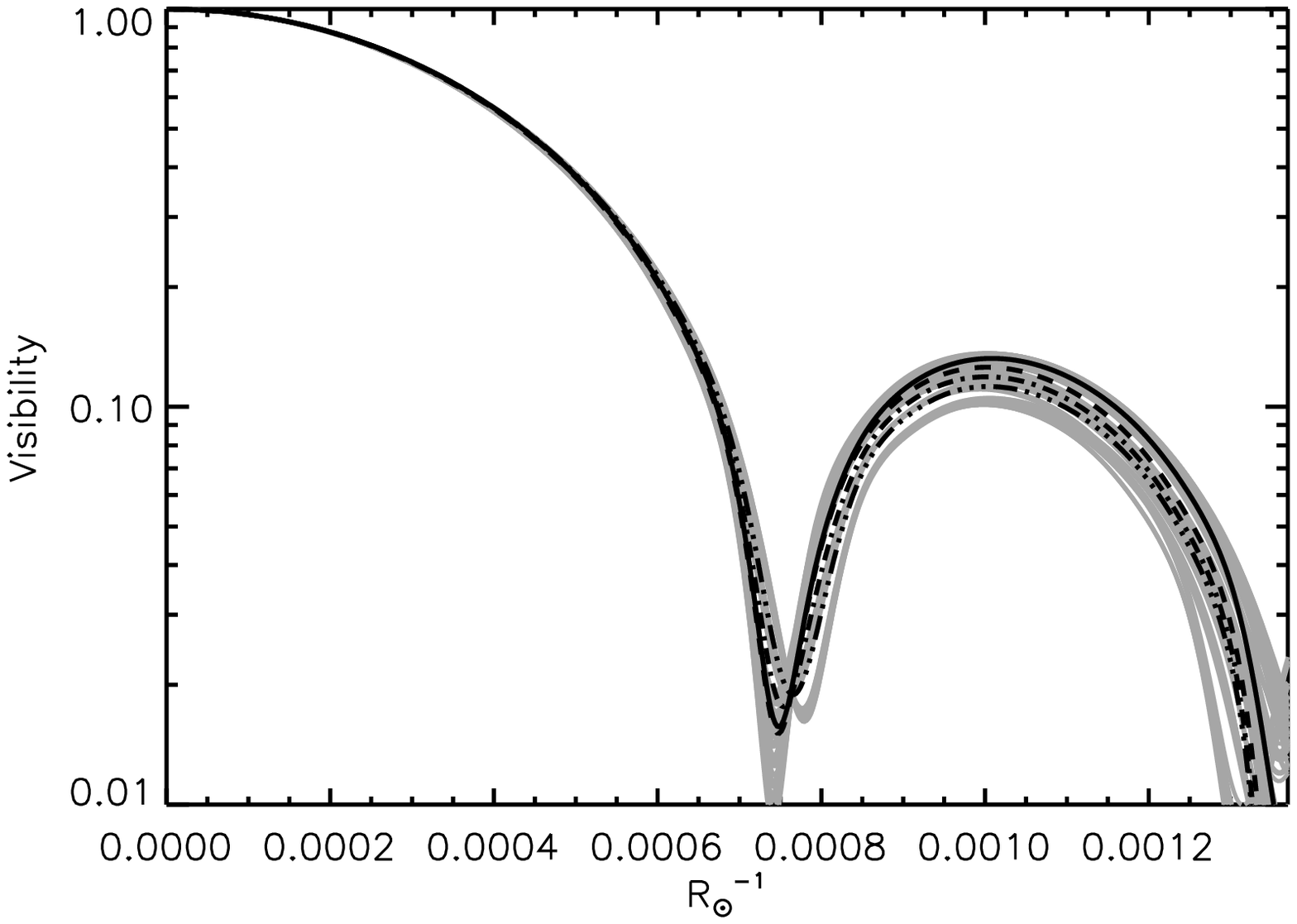} 
 \includegraphics[width=0.5\hsize]{\folder/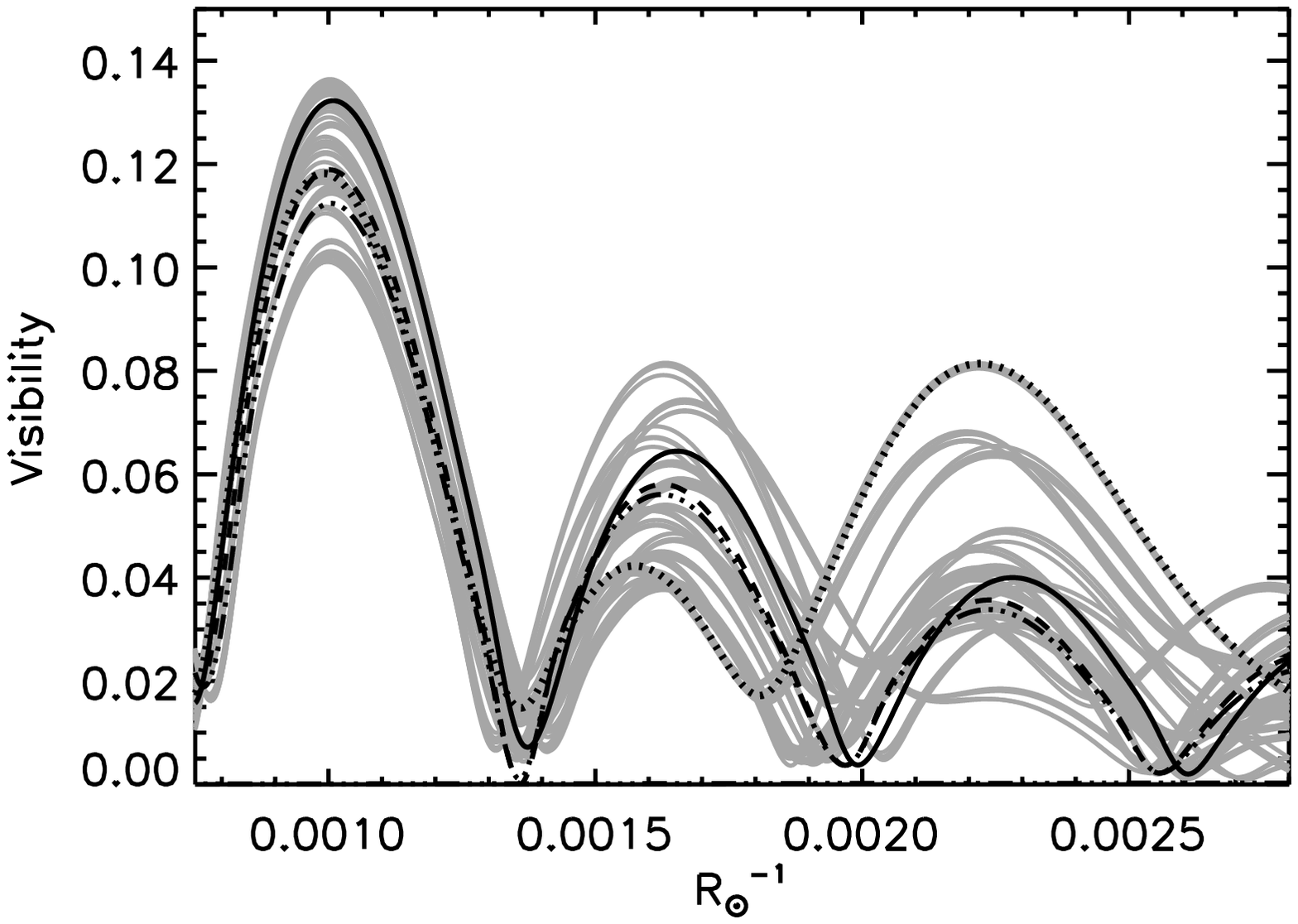}  
 \end{tabular}
      \caption{\emph{Top panel:} intensity maps of a RSG star in the H band. \emph{Bottom left panel:} visibility curves (grey thin line) computed for 36 different angles with a step of 5$^\circ$ a logarithm scale is applied. The solid black curve is a uniform disk model, with a radius of 810 R$_\odot$. The dashed black line is a partially limb-darkened disk ($I\left(\mu\right)=1-0.5\left(1-\mu\right)$) with a radius of 822 R$_\odot$. The dotted dashed line is a fully limb-derkened ($I\left(\mu\right)=1-1\left(1-\mu\right)$) with a radius of 830 R$_\odot$. The stellar parameters of this snapshot are: L = 98400 L$_\odot$, R = 836.5 R$_\odot$, T = 3534 K and log(g) = -0.34. Images from Chiavassa {\em et al.\/}, \cite{2009A&A...506.1351C}.
                       }
         \label{vis_zoom}
   \end{figure}

Fig.~\ref{vis_zoom} (bottom panels) shows the visibility curves
computed for 36 different angles from the intensity map of top
panel. Since we do not know the exact position of the simulated star
with respect to the observations, we applied a statistical approach
and rotated of 5$^\circ$ the intensity map before computing the
$FT$. Small dispersion of the visibility curves (grey thin line)
around the uniform and limb-darkened disks are noticeable: the measure
of the stellar diameter with uniform and limb-darkened disk models
should take into account this dispersion as an additional uncertainty
which may be up to 5$\%$ at the first null point: for example, the
uniform disk spans value between 794 and 845 R$_\odot$ for the
snapshot in Fig.~\ref{vis_zoom} (bottom left panel) in order to fit
the minimum and maximum amplitude of visibility fluctuations
(Chiavassa {\em et al.\/}, \cite{2009A&A...506.1351C}).\\ Moreover,
the visibility dispersion increases clearly with spatial frequency
(bottom right panel), and visibilities strongly deviate from the
parametric cases. It is possible to characterize the typical size
distribution of convective cells on RSGs using interferometric
observables. Chiavassa {\em et al.\/} (\cite{2010A&A...515A..12C})
used alpha Ori (RSG star) observations in H band to tackle this
problem and using the following method: after the computation of the
Fourier transform, $FT$, we obtained
$\hat{I}\left(u,v\right)=FT\left[I\left(x,y\right)\right]$, where
$I\left(u,v\right)$ is a map like in Fig.~\ref{vis_zoom}. The
resulting complex number $\hat{I}\left(u,v\right)$ was multiplied by
low-pass and high-pass filters to extract the information from
different spatial frequency ranges (corresponding to the visibility
lobes). Finally, an inverse Fourier transform, $\bar{FT}$, was used to
obtain the filtered image:
$I_{\rm{filtered}}\left(x,y\right)=\bar{FT}\left[\hat{I}\left(u,v\right)\cdot\rm{filter}\right]$.

\begin{figure}
   \centering
  \begin{tabular}{cc}
\includegraphics[width=0.48\hsize]{\folder/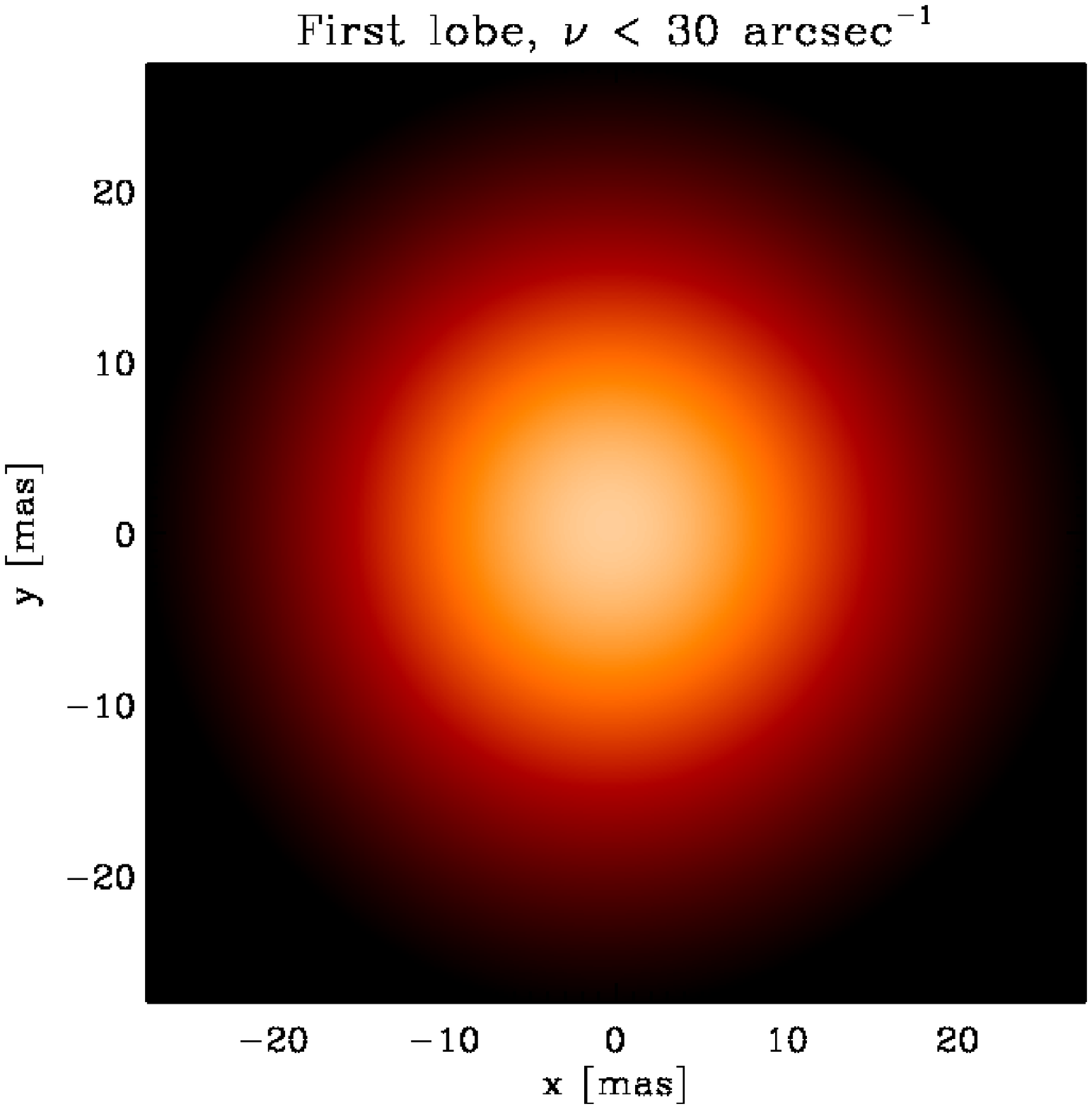}
     \includegraphics[width=0.48\hsize]{\folder/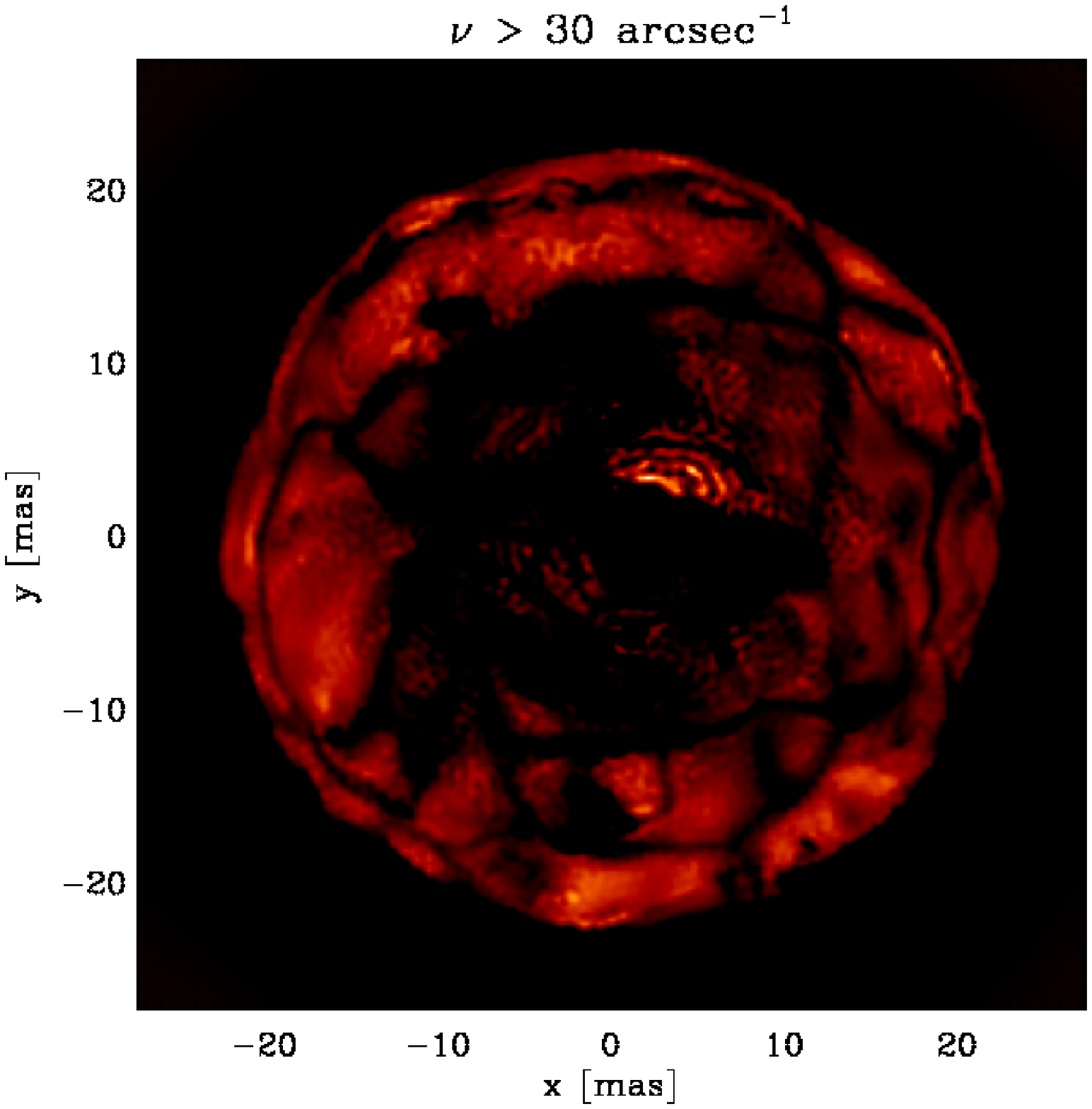}\\
       \includegraphics[width=0.48\hsize]{\folder/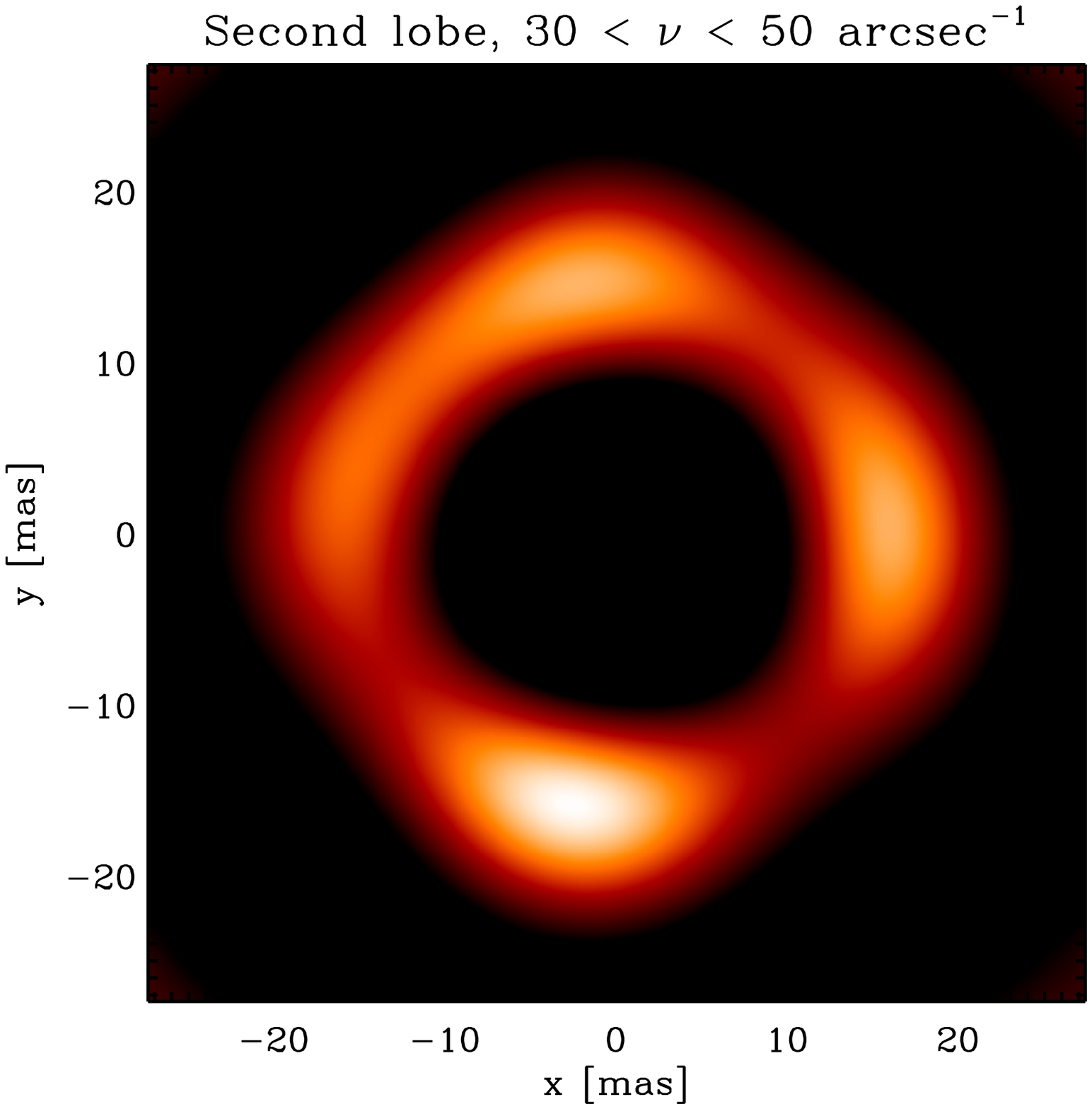}
       \includegraphics[width=0.48\hsize]{\folder/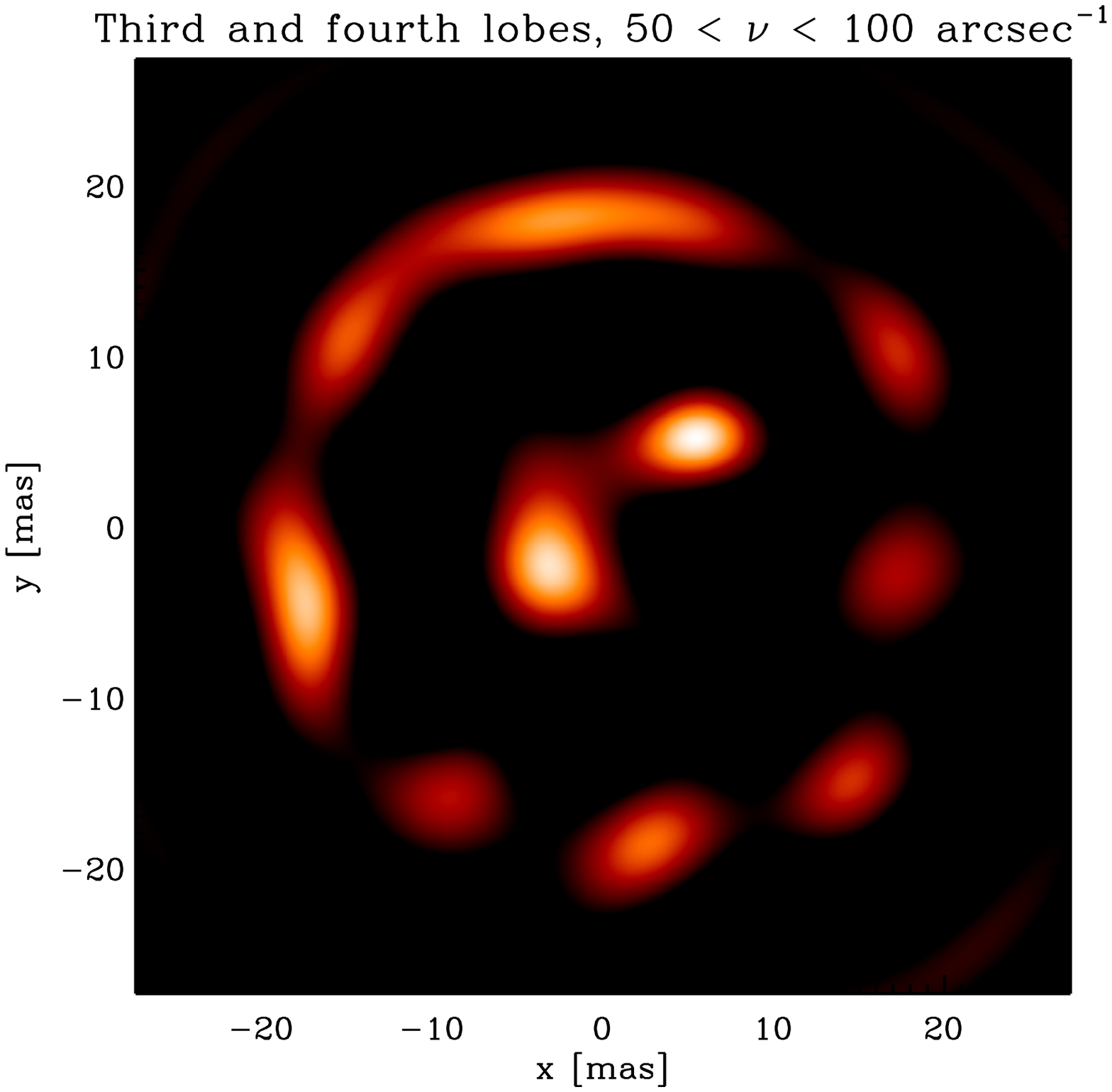}
 \end{tabular}
\caption{Intensity maps filtered at different spatial frequencies
  qualitatively corresponding to the lobes of the visibility curve shown in
  Fig.~\ref{vis_zoom}. The size of alpha Ori is about 45 mas at these wavelengths (H band). Images from Chiavassa {\em et al.\/} (\cite{2010A&A...515A..12C}).}
              \label{size}%
\end{figure}

Figure~\ref{size} (top left panel) shows the filtered images at
spatial frequencies, $\nu$, corresponding to the first lobe of
Fig.~\ref{vis_zoom}. Since we filtered the signal at high spatial
frequencies, the image appears blurry and seems to contain only
information about the stellar radius. However, the top right panel
displays the signal related to all the frequencies higher than the
first lobe: in this image, we clearly do not detect the central
convective cell of $\approx$ 30 mas size ($60\%$ of the stellar
radius, the size of alpha Ori is about 45 mas at this
wavelength). Thus, the first lobe also carries information about the
presence of large convective cells. Figure~\ref{size} (bottom row)
shows the second lobe with convection-related structures of $\approx$
10-15 mas, ($30\%$ of the stellar radius), and the third and fourth
lobes with structures smaller than 10 mas.  We can detect
convection-related structures of different size using visibility
measurements at the appropriate spatial frequencies.

One last important interferometric observables is the closure
phase. The closure phase is defined as the phase of the triple product
(or bispectrum) of the complex visibilities on three baselines, which
form a closed loop joining (at least) three stations A, B, and C. If
the projection of the baseline AB is $\left(u_1,v_1\right)$, that for
BC is $\left(u_2,v_2\right)$, and thus $\left(u_1+u_2,v_1+v_2\right)$
for AC, the closure phase is:

\begin{eqnarray*}
\phi_C (u_1,v_1,u_2,v_2) = 
& \arg ( V(u_1,v_1) & \times V(u_2,v_2) \times V^*(u_1+u_2,v_1+v_2) ) .
\end{eqnarray*}

this procedure removes the atmospheric contribution, leaving the phase
information of the object visibility unaltered
(Monnier, \cite{2007NewAR..51..604M}
and \cite{2003RPPh...66..789M}). Closure phases have the main
advantage of being uncorrupted by telescope-specific phase errors,
including pointing errors, atmospheric piston, and longitudinal
dispersion due to air and water vapor (Le Bouquin {\em et
al.\/}, \cite{2012A&A...541A..89L}).  Eventually, the information
carried by the closure phases is intrinsically correlated with the
morphology of the stellar surface inhomogeneities but depend on the
relative orientation of interferometric baselines and on the
distribution of the telescope station of the diluted aperture of an
interferometer.

\begin{figure}
   \centering
 \includegraphics[width=0.6\hsize]{\folder/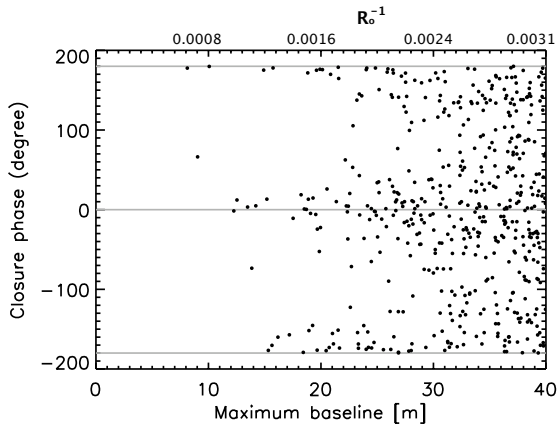}  
      \caption{Scatter plot of closure phases in the H band (from the intensity map of top panel of Fig.~\ref{vis_zoom}) of 500 random 
      baseline triangles with a maximum linear extension of 40 m.
       Closure phases are plotted
      against  the longest baseline of the triangle. 
      The upper x-axis corresponds to synthetic observations of a RSG simulation at an apparent 
      diameter of 43.6~mas (which corresponds to $\alpha$\ Ori at a distance of 174.3\,pc). 
      The axisymmetric case is represented by the grey lines. Image from Chiavassa {\em et al.\/}, \cite{2009A&A...506.1351C}.
}
         \label{clphaFig}
   \end{figure}

Figure~\ref{clphaFig} shows the scatter plot of the closure phase for
a RHD simulation of a RSG in the H band. The closure phases deviate
from zero or $\pm\pi$ (typical values for centrosymmetric objects)
already at $\sim$10 m (0.0008~{\rm R}$^{-1}_\odot$, scaling the
simulation to an apparent diameter of 43.6~mas).  At higher baselines,
it clearly differs from zero or $\pm\pi$, values that are indicating
of a point symmetric brightness distribution.  This is a clear
signature of surface inhomogenities. The characteristic size
distribution on the stellar surface can also be derived from the
closure phase: the contribution of small-scale convection-related
surface structures increases with frequency.  It may be very efficient
to constrain the level of asymmetry of RSG atmospheres by accumulating
statistics on closure phase at short and long baselines, since they
are easily measured to high precision. A small departure from zero
immediately inters a departure from symmetry (Chiavassa {\em et
al.\/}, \cite{2009A&A...506.1351C}).

RHD simulations are necessary for a proper quantitative analysis of
interferometric observations of the surface of evolved stars beyond
the smooth, symmetrical, limb-darkened intensity profiles. Frequencies
corresponding to the first lobe (ie., sensitive to the radius) are
characterized by fluctuations as high as 5$\%$, and radii
determinations can be affected to that extent. The second, third,
fourth lobes, and so on carry the signature of limb-darkening, and of
smaller scale structures due the granulation, and are very different
from the simple parametric cases.  Also closure phases largely differ
from 0 and $\pm\pi$ with an increase of the signal with frequency,
this is due to the departure from circular symmetry. The
characteristic size of the granulation pattern of, eg., RSGs is
composed of convection-related structures of different sizes,
including small to medium scale granules (5--15 mas) and a large
convective cell ($\approx$30 mas).

\begin{figure}
   \centering
   \includegraphics[width=0.6\hsize]{\folder/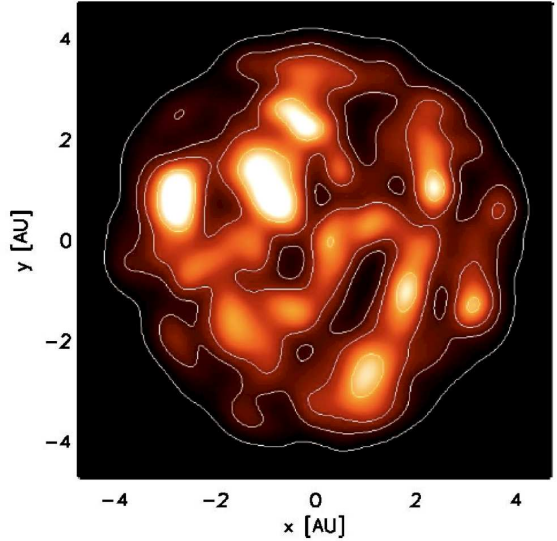}
      \caption{Snapshot of a RSG simulation convolved to the PSF of a 42\,m
        telescope approximatively like the European Extremely Large
        Telescope (for a star at a distance of 152.4 pc). Image from Chiavassa {\em et al.\/}, \cite{2011A&A...528A.120C}.}
         \label{ELT}
   \end{figure}

In the last years, RHD simulations managed to explain several
interferometric observations of RSGs (Chiavassa {\em et
al.\/}, \cite{2009A&A...506.1351C}, \cite{2010A&A...515A..12C}) and
AGBs (Chiavassa {\em et al.\/}, \cite{2010A&A...511A..51C}) from the
visible to the infrared region. However, the final consistency check
will be an image reconstruction to directly compare the granulation
size and shape and the intensity contrast, provided by the planned
second generation recombiner of the VLTI (MATISSE, GRAVITY, see other
contributions is this book) and CHARA optical interferometry
arrays. First steps in this direction has been carried out in Berger
{\em et al.\/} (\cite{2012A&ARv..20...53B}) and Malbet {\em et al.\/}
(\cite{2010SPIE.7734E..83M}) where the image reconstruction algorithms
have been tested using intensity maps from these RHD simulations. The
European Extremely Large Telescope (E-ELT, planned to be operating in
2025) with a mirror size five times larger than a single VLT Unit
Telescope will be capable of near IR observations of surface details
on RSGs (Fig.~\ref{ELT}).

    \section[From Main Sequence to evolved stars]{Transition from main sequence stars to evolved stars}

\begin{figure*}
   \centering
   \begin{tabular}{cc}
                           \includegraphics[width=1.\hsize]{\folder/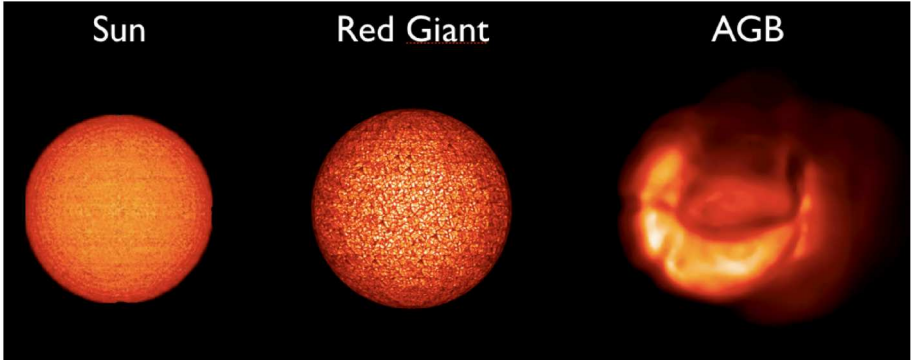}\\
 	           \end{tabular}
      \caption{Evolution of the surface inhomogeneities during the stellar evolution: main sequence (\emph{left panel}), red giant (\emph{central panel}), and AGB (\emph{right panel}) (the scale is not preserved).}
        \label{evolution}
   \end{figure*}

The surface inhomogeneities at the surface of the star, i.e. the
granulation aspect, depend on the effective temperature and even more
important the surface gravity.  The size of the granule is controlled
by the stratification and the exponential drop of density at the
stellar surface (e.g. Nordlund {\em et
al.\/} \cite{1990A&A...228..155N} and Stein {\em et
al.\/} \cite{1998ApJ...499..914S}). The effective temperature and
metallicity play a less important role (Magic and
Asplund \cite{2014arXiv1405.7628M}). On the Sun, the average granule
size is about 1-2 Mm, therefore there is at least $10^{7}$ granules on
the entire surface. The consequence is that even if the granule aspect
is very irregular, their tiny size with respect to the stellar radius
makes the appearance of the surface homogenous
(Fig.~\ref{evolution}). During the evolution of a solar-type star, the
radius increases by more than an order of magnitude which lower the
surface gravity, e.g. ${\rm log\,g}\sim2$ for the red giant phase and
even lower than 0 for the AGB phase. The star inflates but at the same
time the granule size increases to by a factor of $10^3$ (Magic and
Asplund \cite{2014arXiv1405.7628M}). The number of granules covering
the entire surface for a red giant is much smaller than on the main
sequence phase, with roughly a few thousand : the stellar surface
appears more and more irregular as the star evolves.

Clearly, for the latest stage of the stellar evolution the surface of
the star is very inhomogenous. The operation of computing many Fourier
transform (FT) is however costly from the numerical point of view. In
the process of deriving the angular diameter of the star, one needs to
compute many time the visibilities of the fringes using for example
Monte Carlo procedure to derive reliable diameter and its error bars.
Most of the objects for which we need precise angular diameters are
main sequence stars or early phase of red giants for which we
can combine asteroseismology and interferometric determination of the
diameter to tidily constrain the other stellar parameters (e.g.,
Creevey this book; Bigot {\em et al.\/} \cite{2011A&A...534L...3B}
and \cite{bigot15}). In these cases we can take
advantage of the relatively homogenous surface. The idea is to
simplify the FT using axial symmetry of the star. The steps are
described below.

The usual visibility in the $(u,v)$-plane can be computed using the
theorem van Citter-Zernike (e.g., Born and
Wolf \cite{1999prop.book.....B})

\begin{equation}
F(u,v)= FT \left [I(x,y) \right ] = \int I(x,y) e^{2\pi i(x.u+y.v)} dx dy.
\end{equation}

This 2D Fourier transform can be written in polar coordinates
$(r,\theta)$ where $r$ is the position of the stellar disk $\theta$
the azimuthal angle. Therefore $x=r\cos\varphi$ and $y=r\sin\varphi$
and the surface element can be transformed as $dxdy =r drd\varphi$
($r$ being the Jacobian of the transformation). The Fourier transforms
writes

\begin{equation}
F(u,v)= \int  I(r,\varphi) e^{2\pi ir(u\cos\varphi+v\sin\theta)} rdrd\varphi.
\end{equation}

It is convenient to write the coordinate $(u,v)$ also in polar coordinates $(k,\phi)$. Let us, therefore define the following $u=k\cos\phi$ and $v=k\sin\phi$. The Fourier transform then writes  
\begin{figure*}
   \centering
   \begin{tabular}{cc}
                           \includegraphics[width=0.99\hsize]{\folder/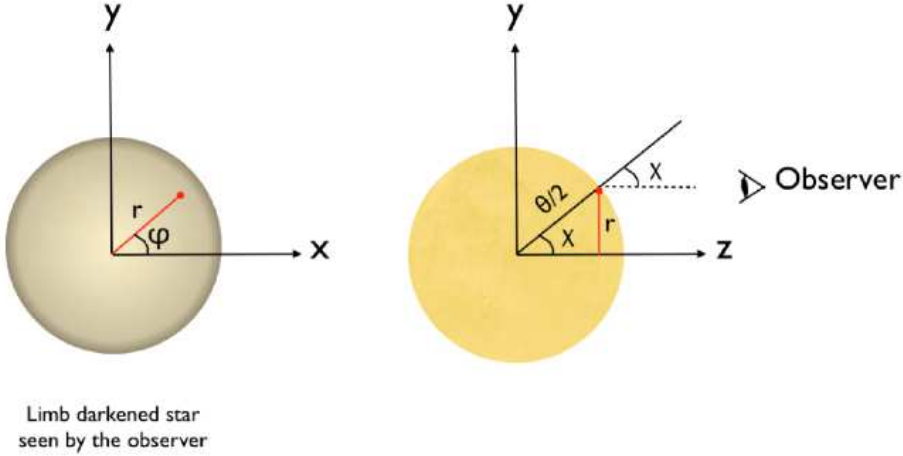}\\
 	           \end{tabular}
      \caption{\emph{Left panel: }Geometry of the system. the star with axial symmetry is represented. The surface brightness depends on the position on the stellar disk, $r$, because of the kimb darkening, but does not depend on $\varphi$ as long as the granules are very small compared with the radius. \emph{Right panel:} Cut of the star in the plane perpendicular to the left panel. The $\chi$ angle measures the position on the stellar disk}
        \label{schema}
   \end{figure*}

\begin{equation}\label{fourier_step1}
F(k,\phi)= \int I(r,\varphi) e^{2\pi irk(\cos\phi\cos\varphi+\sin\phi\sin\varphi)} rdrd\theta =  \int I(r,\varphi) e^{2\pi irk \cos(\varphi-\phi)} rdrd\varphi.
\end{equation}

It is important to note that no assumption is made to derive
Eq. \ref{fourier_step1}. If the intensity is now axisymmetric then $
I(r,\varphi) = I(r)$ and we can separate the radial and angular
integration. The latter is identical to a zeroth-order Bessel function
of the first kind (Abramowitch and Stegun\cite{1972hmfw.book.....A}).

\begin{equation}
\int e^{ikr\cos(\varphi-\phi)} d\theta  = 2\pi J_0(kr).
\end{equation}

In the latter expression, the $\phi$ coordinate disappeared since for
any value of $\phi$ we can always make a change of variable so that
$\int e^{2\pi ikr\cos(\varphi-\phi)} d\varphi=\int e^{2\pi
ikr\cos(\varphi')} d\varphi'$.  The Fourier transform of the stellar
disk intensity in the case of axial symmetry simplifies to a 1D
integral

\begin{equation}
F(k) = 2\pi \int_0^{\theta/2} I(r) J_0(kr) rdr,
\end{equation}

which is the Hankel transform of the disk intensity $I(r)$. Since $k$
is the spatial frequency we have $k=B/\lambda$, where $B$ is the
baseline between the two telescopes and $\lambda$ is the wavelength of
the measurement. The position, $r$, on the stellar disk measures the
distance between the disk center $r=0$ and the limb $r=\theta/2$,
where $\theta$ is the angular diameter of the star. It is common to
express since distance is term of the angle $\chi$ made by the
vertical (disk center) and the line-of-sight of an observer (see
Fig.~\ref{schema}). Therefore the position of the stellar disk is
simply $r=\theta/2 \sqrt{(1-\mu^2)}$ and $dr/r = -1/(1-\mu^2)\mu d\mu$
where $\mu=\cos\chi$.

With these definition we car rewrite the  visibilities as
\begin{equation}
V \propto \int_0^1 I(\mu) J_0(\pi\sqrt{1-\mu^2}\theta B/\lambda) \mu d\mu
\end{equation}
which is the usual expression using for fitting angular diameters from the computed limb-darkened intensities. 
   
   \subsection{Main sequence and giant stars}

\begin{figure}
   \centering
   \begin{tabular}{c}
        \includegraphics[width=0.98\hsize]{\folder/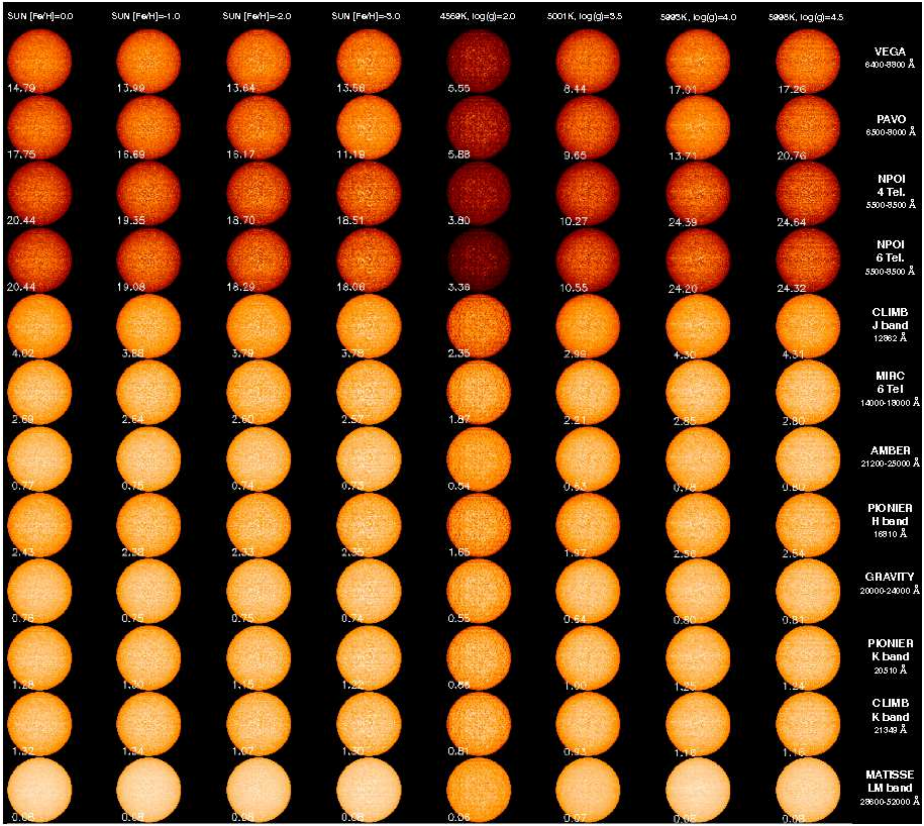}\\
  
          \end{tabular}
      \caption{Synthetic stellar disk images of different RHD simulations (columns). The images correspond to a representative wavelength for each interferometric instruments from the visible (top row) to the far infrared (bottom row). The averaged intensity ($\times10^5$\,erg\,cm$^{-2}$\,s$^{-1}$\,{\AA}$^{-1}$) is reported in the lower left corner of each image. Image from Chiavassa {\em et al.\/} (\cite{2014A&A...567A.115C}).}
        \label{images}
   \end{figure}

As explained in Section~\ref{Sectmodel}, the computational domain of
RHD simulations of main sequence and K giant stars represents only a
small portion of the stellar surface. To overcome this limitation, and
at the same time account for limb-darkening effects, Chiavassa {\em et
al.\/}
(\cite{2010A&A...524A..93C}, \cite{2012A&A...540A...5C}, \cite{2014A&A...567A.115C})
computed intensity maps for different inclinations ($\theta$-angles)
with respect to the vertical direction and for representative series
(covering several convection turnover) of simulation snapshots and
used them to tile a spherical surface. The computed values of the
$\theta$-angle depend on the position (longitude and latitude) of the
tile on the sphere and are linearly interpolated among the inclination
angles. In addition to this, the statistical tile-to-tile fluctuations
(i.e., number of granules, shape, and size) is taken in consideration
by selecting random snapshots within each simulation's time-series. As
a consequence, the simulation assumption of periodic boundary
conditions resulted in a tiled spherical surface globally displaying
an artifactual periodic granulation pattern. However, the resulting
artificial signal introduced in the interferometric observables is
less important that the one caused by the inhomogeneities of the
stellar surface. The final result is an orthographic projection (i.e.,
stellar disk image as seen from a nearby observer) of the tiled
spheres (Fig.~\ref{granulation} and \ref{images}).

The extraction of interfererometric observables is done using the
stellar disk images and the $FT$ method described in the previous
sections. However, only closure phases are sensitive enough to the
effect of stellar granulation to expect a possible detection with
today interferometers. Chiavassa {\em et al.\/}
(\cite{2014A&A...567A.115C}) computed the closure phases for twelve
interferometric instruments covering wavelengths ranging from visible
to infrared, optimizing the observability to allow a broad coverage of
spatial frequencies. The aim of that work is to present a survey of
the granulation pattern of stars with different stellar parameters and
to evaluate its effect on the detection of planet
transit. Fig.~\ref{images} displays inhomogeneous brightness
distributions in the disk images; moreover, the centre-to-limb
variations are more pronounced in the visible instruments with respect
to the infrared ones. This effect is explained by different
sensitivity of the source (Planck) function at visible and at infrared
wavelengths.

  \begin{figure}
   \centering
   \begin{tabular}{c}
              \includegraphics[width=0.5\hsize]{\folder/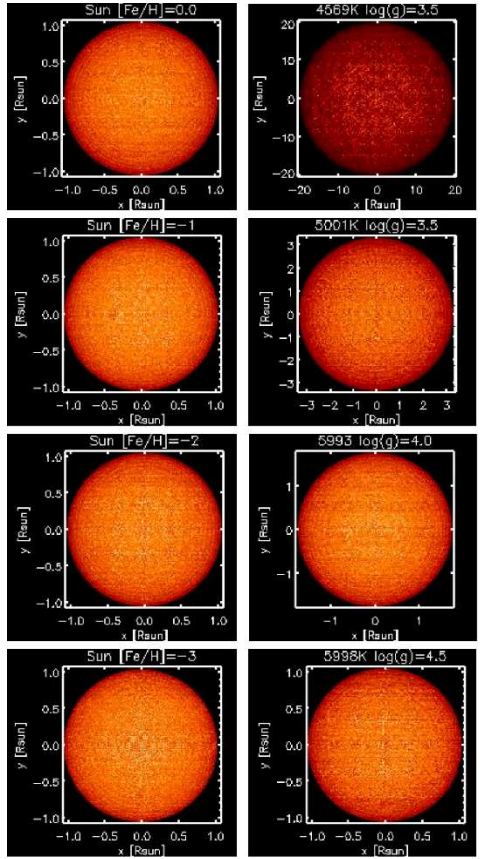}
        \end{tabular}
      \caption{Enlargement of the synthetic stellar disk images of Fig.~\ref{images} for the VEGA instrument mounted at CHARA interferometer. Image from Chiavassa {\em et al.\/} (\cite{2014A&A...567A.115C}).}
              \label{imagesbis}
   \end{figure}
   
   \begin{figure*}
   \centering
   \begin{tabular}{cc}
                           \includegraphics[width=0.65\hsize]{\folder/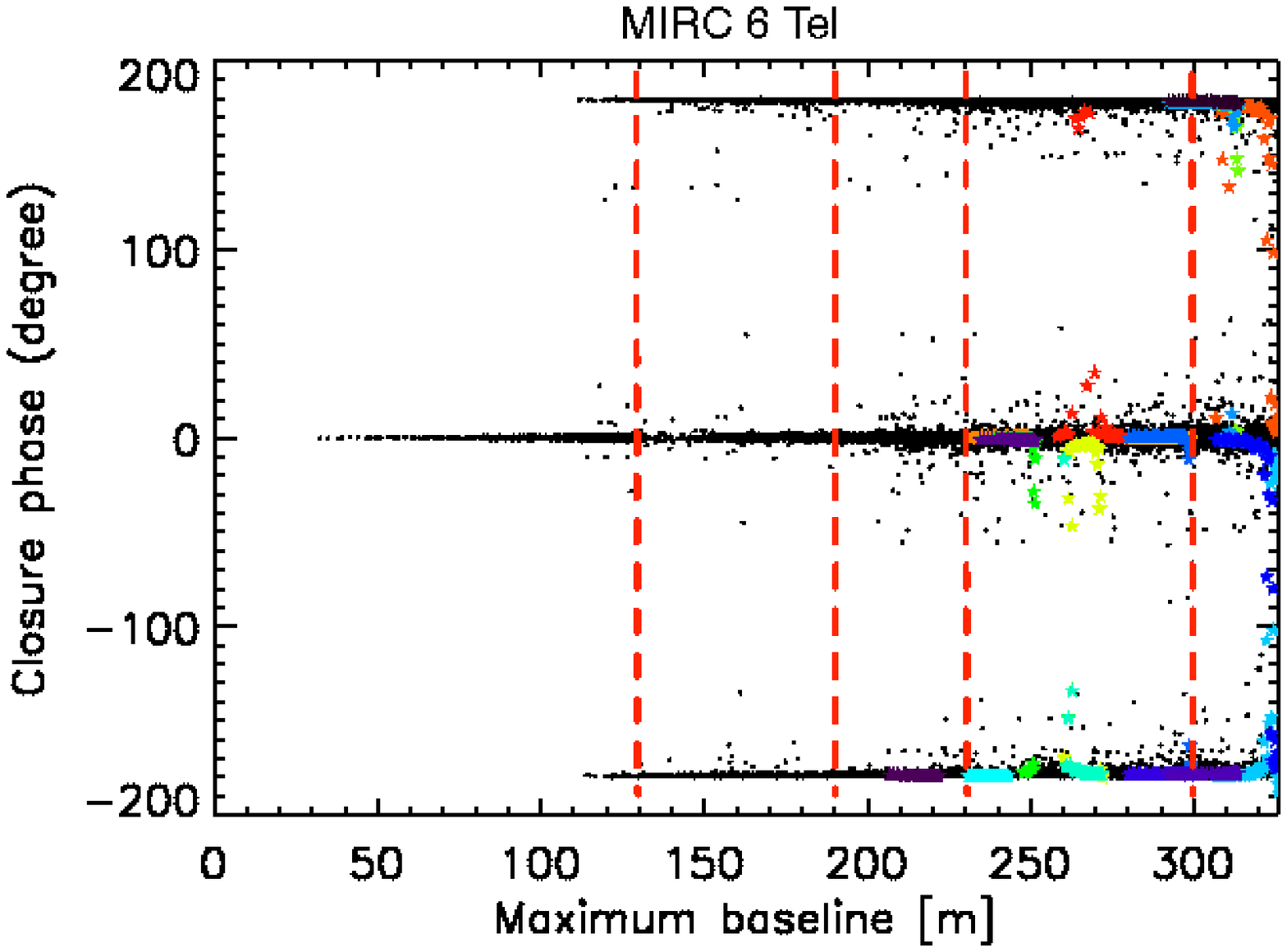}\\
                            \includegraphics[width=0.7\hsize]{\folder/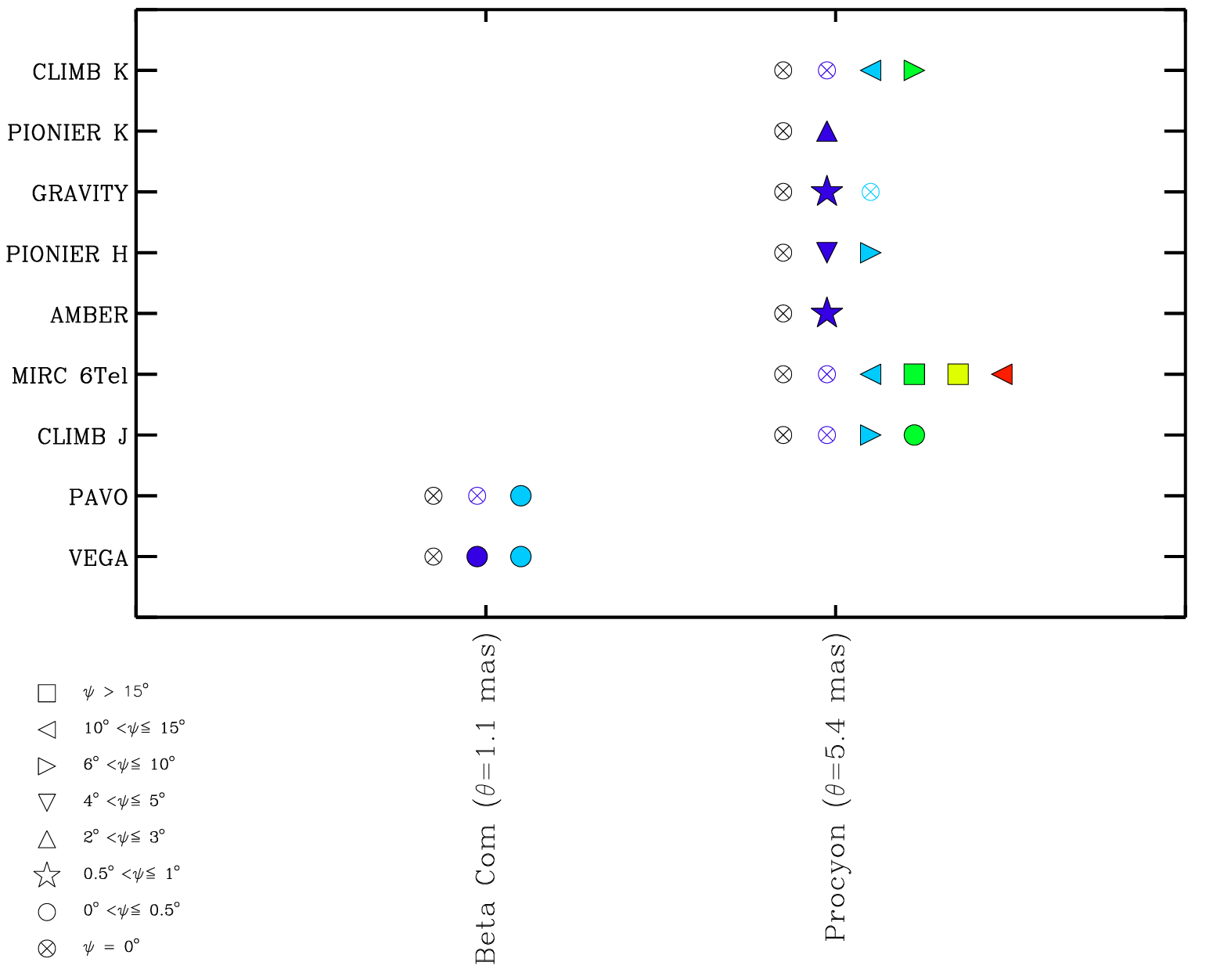}
 	           \end{tabular}
      \caption{\emph{Top panel:} Scatter plot of closure phases of 20000 random baseline triangles (black dots) as a function of the maximum linear extension corresponding to the configuration chosen for each MIRC instrument and for the RHD simulation of the Sun. The colored symbols over-plotted display the closure phases for a particular telescope configuration chosen and the vertical dashed red lines give the approximative positions of the different lobes. \emph{Bottom panel:} Departure, $\psi$, of closure phases from zero and/or $\pm\pi$ (i.e., no axisymmetric case) for RHD simulations with stellar parameters corresponding to two real stars (horizontal axis) and different interferometric instruments (vertical axis). 6 lobes are displayed: black for the 1st lobe, violet for the 2nd, light blue for the 3rd, green for the 4th, yellow for the 5th, and red for the 6th. Only the lobes spanned in the UV-planes are plotted. The symbols correspond to different values, in degrees. Images from Chiavassa {\em et al.\/} (\cite{2014A&A...567A.115C}).}
        \label{closure}
   \end{figure*}

Figure~\ref{closure} (top panel) displays one example from Chiavassa
{\em et al.\/} (\cite{2014A&A...567A.115C}) of closure phases
deviating from the axisymmetric case for MIRC-6 telescope
configuration case. The scatter plot of closure phase increase
particularly in the visible wavelength range, where the dispersion is
larger (e.g., VEGA, NPOI, and PAVO instruments). Depending on the
instruments and spatial frequency spanned, the departures from
symmetry may be large or not. However, it is apparent that the
convection-related surface structures have a signature on the closure
phases.

Figure~\ref{closure} (bottom panel) shows indeed a direct application
to two real targets: Beta Com, G0V star with angular diameter of 1.1
mas (Richichi {\em et al.\/}, \cite{2005A&A...431..773R}); and
Procyon, F5IV with angular diameter of 5.4 mas (Chiavassa {\em et
al.\/}, \cite{2012A&A...540A...5C}). In general, all the instruments
(except for MATISSE and NPOI, which do not probe frequencies larger
than the first lobe) show closure phases departures ($\psi$) of the
order of few degrees with largest values of the order of
$\sim16^\circ$. These values are much larger than what are the actual
instrument's incertitudes. More in detail:

\begin{itemize}
\item PAVO and VEGA show departures  lower than 0.5$^\circ$ already on the 2nd lobe (VEGA instrument);
\item AMBER and GRAVITY with values lower than 0.8$^\circ$;
\item PIONIER in the H band with values of 4.3$^\circ$ (2nd lobe) 6.4$^\circ$ (3rd lobe), and PIONIER in the K band with 2.9$^\circ$ (2nd lobe);
\item MIRC with 13.8$^\circ$, 15.3$^\circ$, 16.4$^\circ$, 13.1$^\circ$ for the 3rd, 4th, 5th, 6th lobe, respectively;
\item CLIMB J band with 6.5$^\circ$, 0.2$^\circ$, 5.1$^\circ$ for the 3rd, 4th, 5th lobe, respectively;
\item CLIMB K band with 12.3$^\circ$, 6.3$^\circ$ for the 3rd, 4th lobe, respectively.
\end{itemize}

The actual instruments and telescopes allow, in principle with very
good weather conditions, the detection of the granulation. The closure
phase signal is already more pronounced in the infrared for the 2nd
lobe and may be detected with very good weather and instrumental
conditions but it is certainly easier to detect from the 3rd lobe
on. MIRC instrument with 6 telescope recombination is the most
appropriate instrument as it combines good UV coverage and long
baselines probed.

   \begin{figure}
   \centering
 \begin{tabular}{ccc}
	     \includegraphics[width=0.32\hsize]{\folder/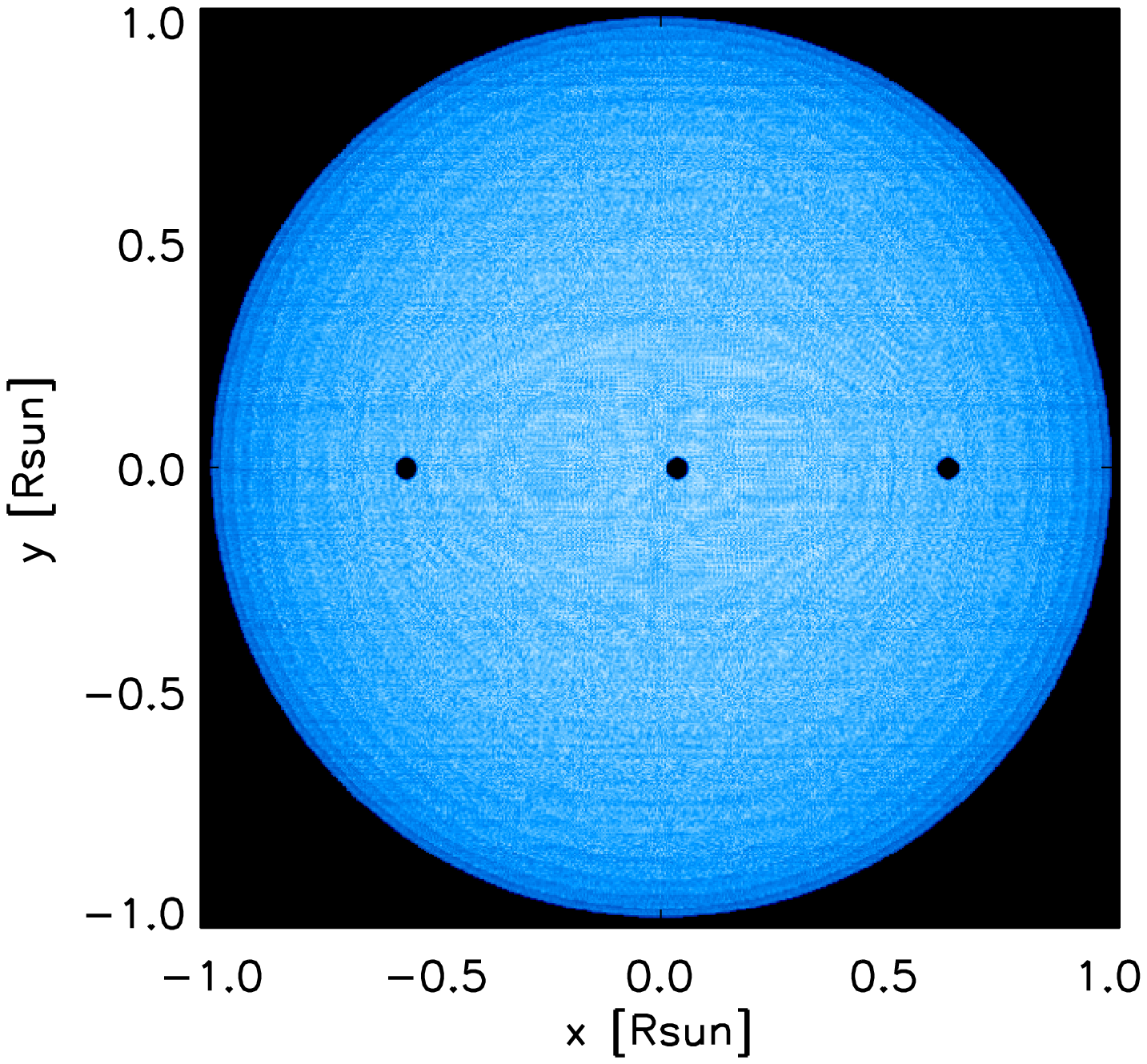}
              \includegraphics[width=0.32\hsize]{\folder/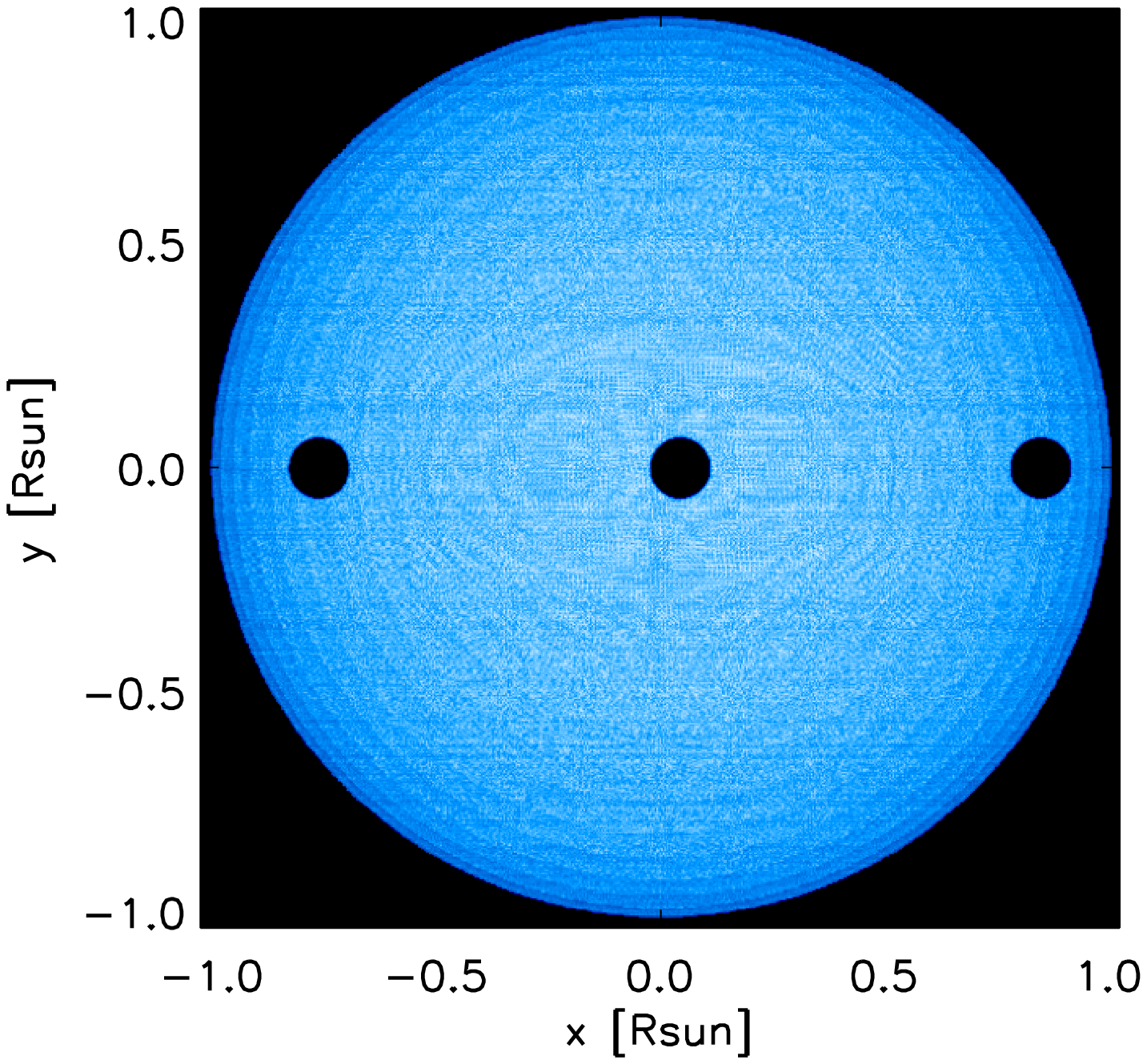}
              \includegraphics[width=0.32\hsize]{\folder/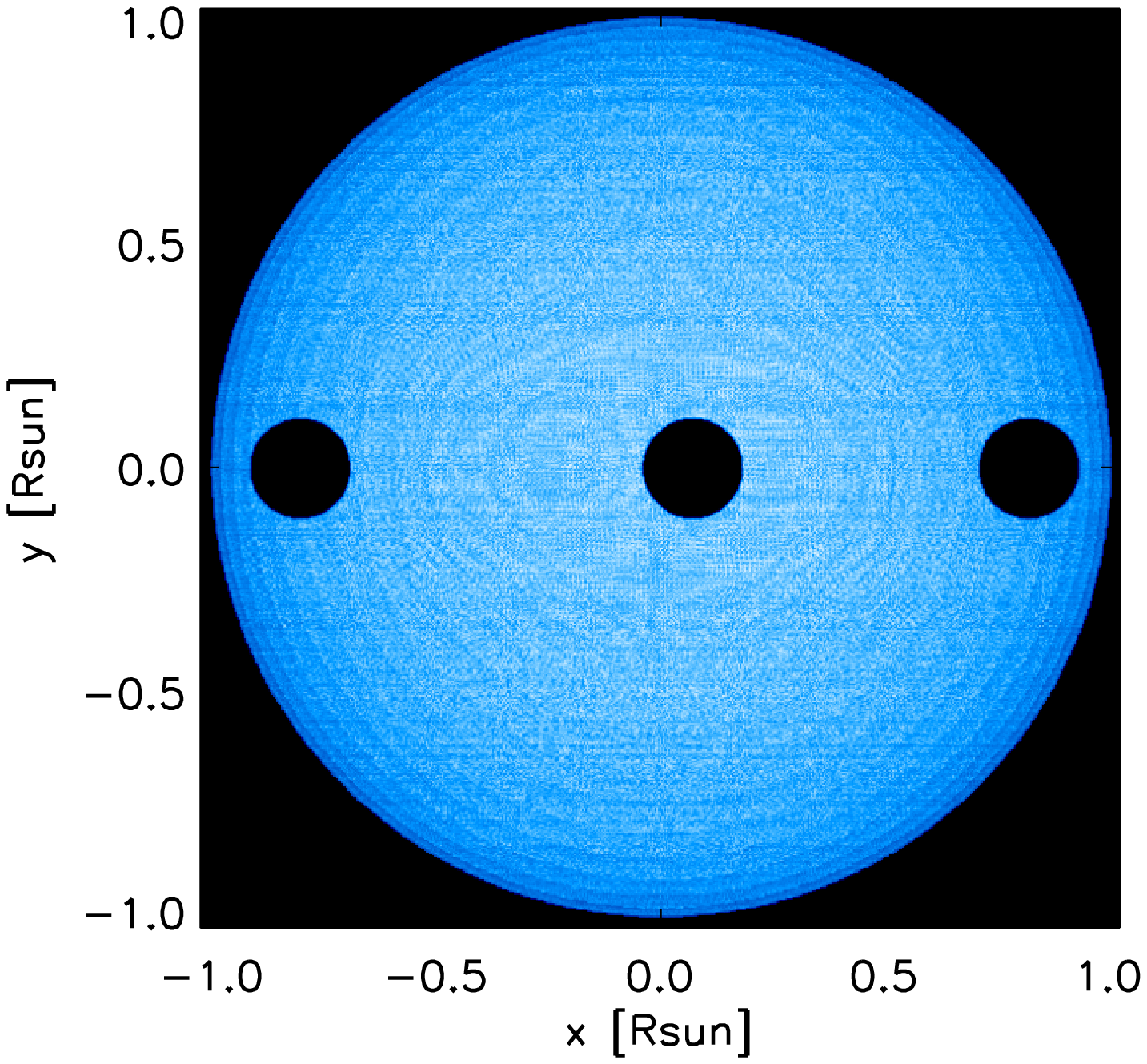}\\
              \includegraphics[width=0.7\hsize]{\folder/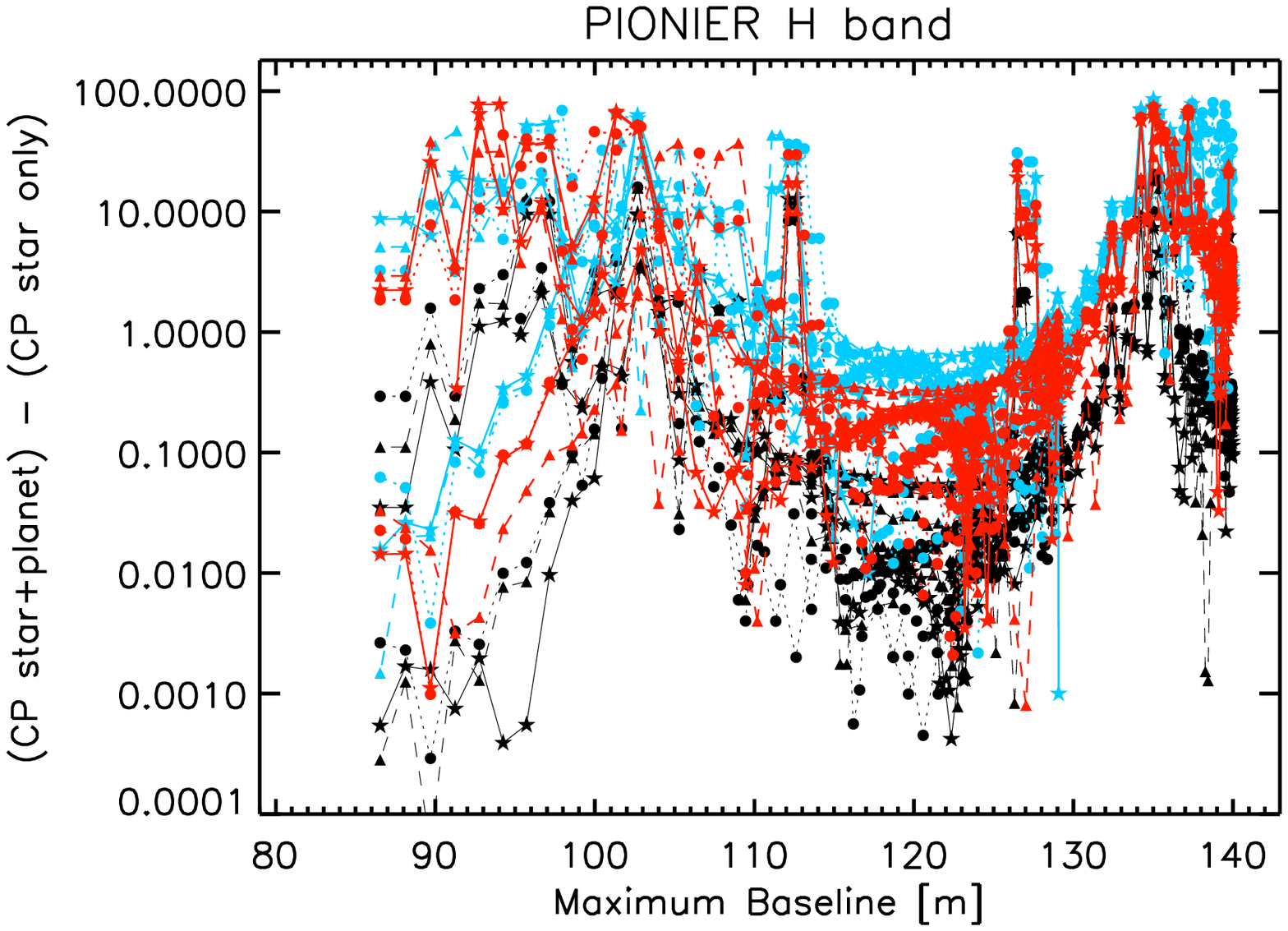}
         \end{tabular}
      \caption{\emph{Top:} Synthetic stellar disk images in H band (PIONIER instrument) together with three planet transiting phases (black color) for a star like the Sun. The prototypes of planet have parameters corresponding to: Kepler-11 f (left), HD 149026 b (central), and CoRoT-14 b (right). \emph{Bottom:} Absolute closure phase differences (in degrees) between the star with a transiting planet (from above) and the star alone (Fig.~\ref{images}). The black colour correspond to Kepler 11-f, the red to HD 149026 b, and the blue CoRoT 14-b. The star symbols connected with solid lines correspond to the planet phase entering in the stellar disk, the circle symbols connected with dotted line to the planet at the centre of the stellar disk, and the triangles connected with dashed line to the planet exiting the stellar disk. Images from Chiavassa {\em et al.\/} (\cite{2014A&A...567A.115C}).}
        \label{transit}
   \end{figure}

Another important role of long-baseline interferometric observations
is planet hunting as a complement to the radial velocity and adaptive
optics surveys. High angular resolution is an ideal tool for exploring
separations in the range 1 to 50 mas (Le Bouquin {\em et
al.\/}, \cite{2012A&A...541A..89L}). In addition to this, Chiavassa
{\em et al.\/} (\cite{2014A&A...567A.115C}) measured the contamination
of granulation signal on the planetary transit using closure
phases. To do this, the authors chose three prototypes of planets
representing different sizes and compositions because the purpose is
not to reproduce the exact conditions of the planet-star system
already detected but to have a statistical approach on the
interferometric signature for different stellar parameters hosting
planets with different sizes. Then they created transiting images
(Fig.~\ref{transit}, top) and computed the resulting closure
phases. Eventually, they determined the differences between
planet-star system and the star alone (Fig.~\ref{transit},
bottom). For all the instruments, the absolute difference scales with
the size of the planet considered: the smaller planet returns smaller
differences. In addition to this, the differences are larger in the
visible wavelengths where the granulation contrast is higher than in
the infrared.

The signature of the transiting planet on the closure phase is mixed
with the signal due to the convection-related surface structures. The
time-scale of granulation depends on the stellar parameters, and
varies from minutes or tens of minutes for solar type stars and
sub-giants, to hours for more evolved red giant stars. If the transit
is longer that the granulation time-scale (which is the case for most
of main sequence stars), it is possible to disentangle its signal from
convection by observing at particular wavelengths (either in the
infrared or in the visible) and measuring the closure phases for the
star at difference phases of the planetary transit.
  
For this purpose, it is very important to have a comprehensive
knowledge of the host star to detect and characterize the orbiting
planet, and RHD simulations are very important to reach this aim.

\section{Stellar parameters}

\begin{figure*}
  \centering
   \begin{tabular}{cc}
 \includegraphics[width=0.5\hsize]{\folder/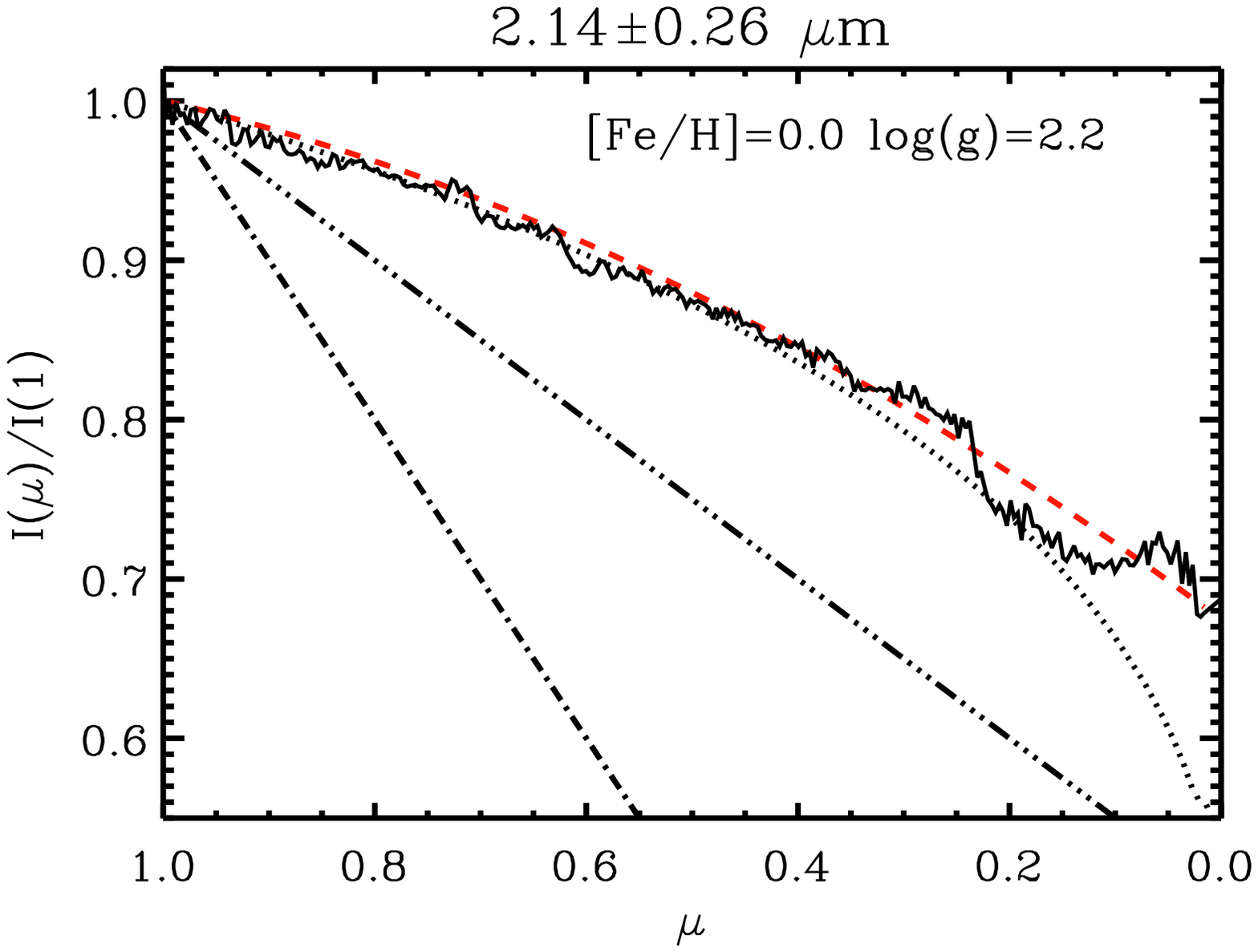} 
 \includegraphics[width=0.5\hsize]{\folder/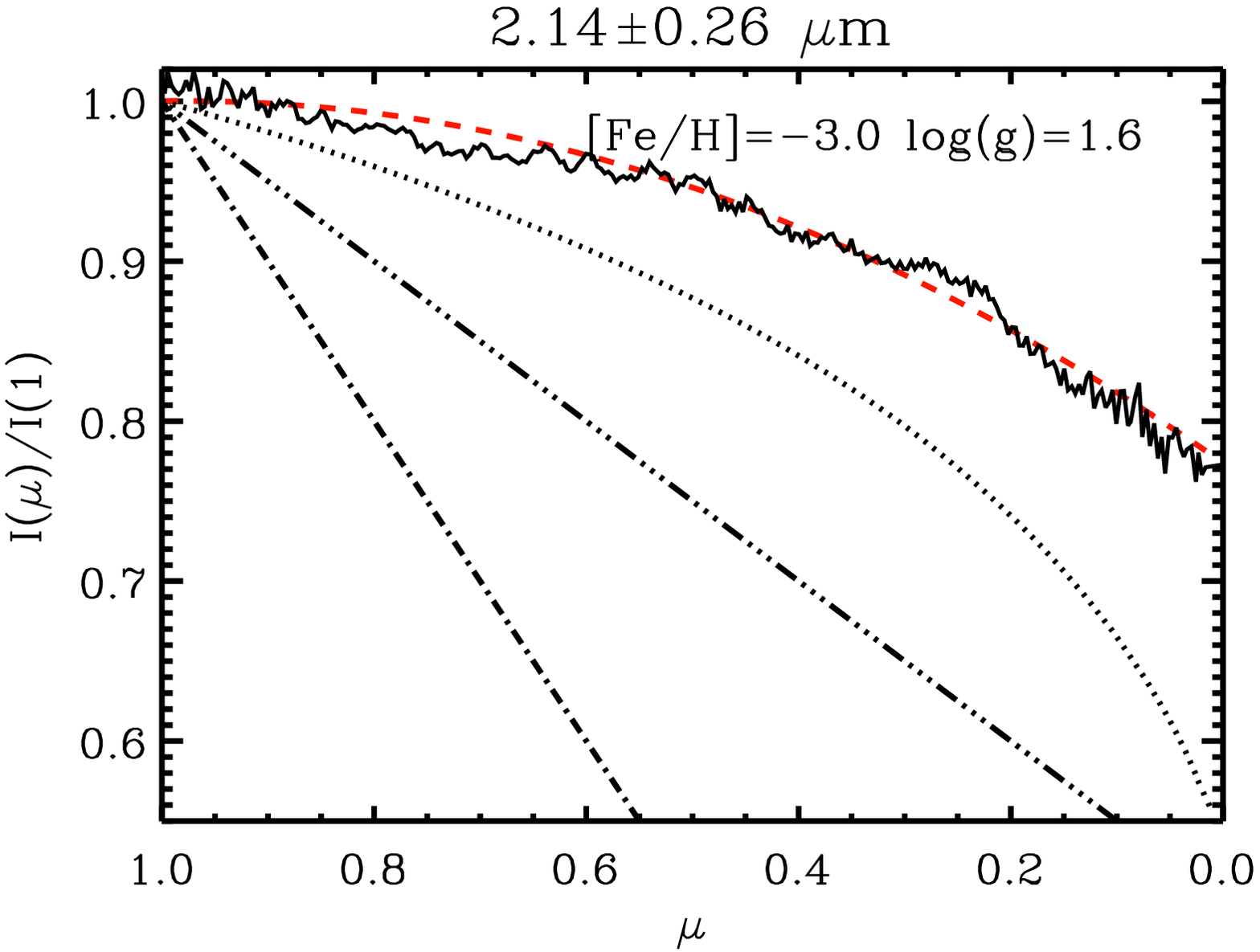} \\
  \includegraphics[width=0.5\hsize]{\folder/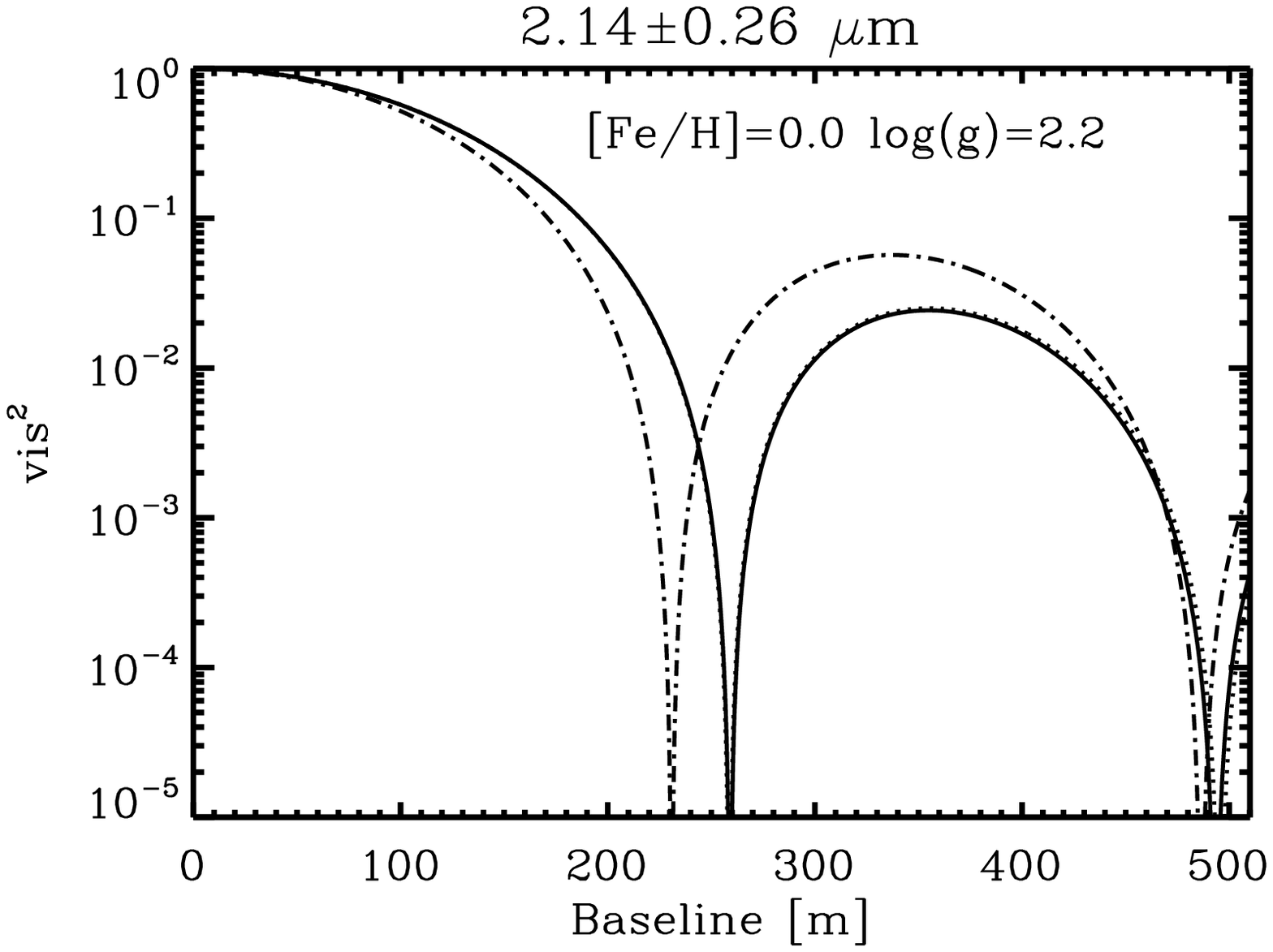}
   \includegraphics[width=0.5\hsize]{\folder/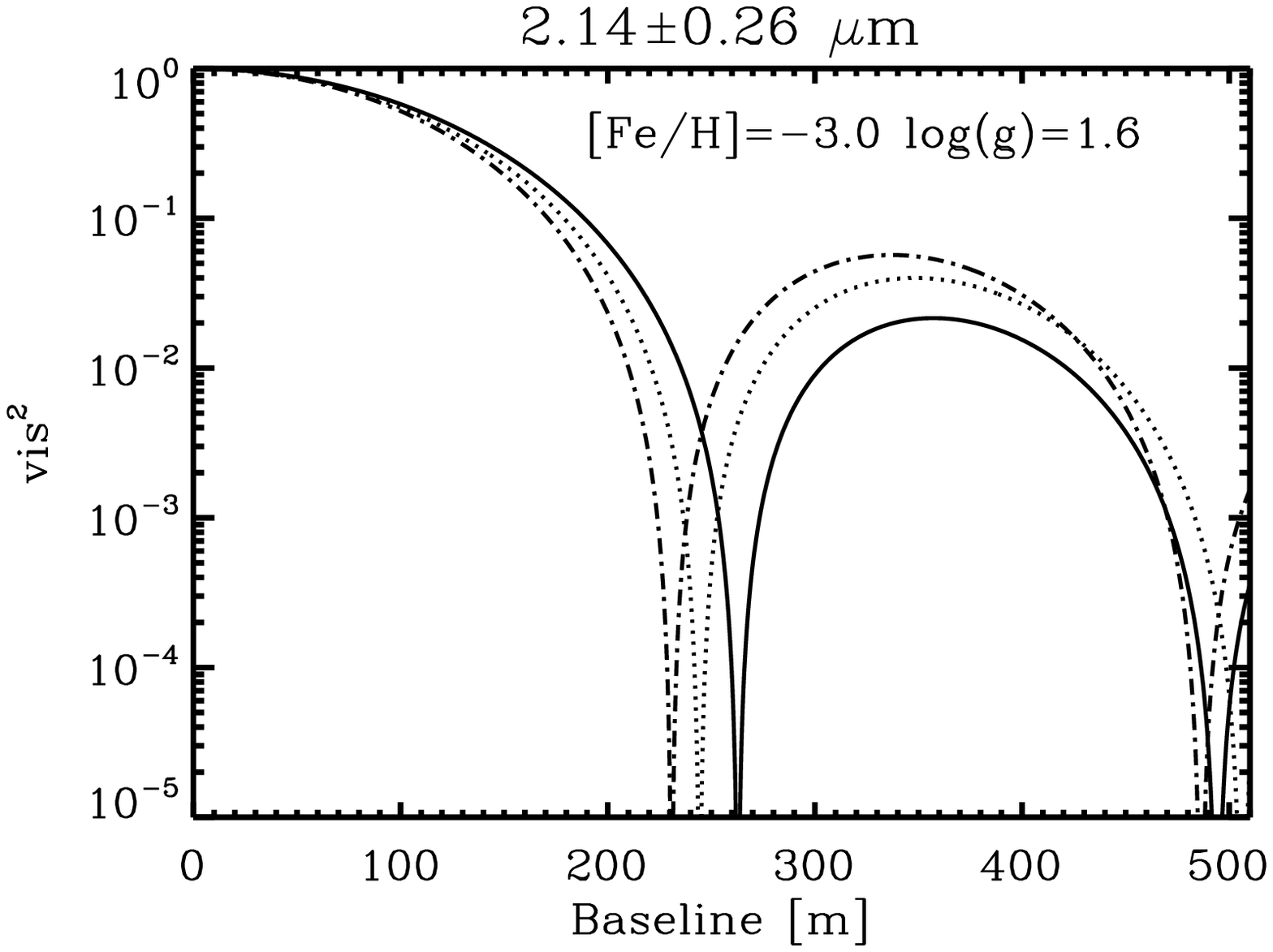} \\
 \end{tabular}
     \caption{\emph{Top panels:} limb-darkening fits (red dashed line) for the RHD azimuthally average
       intensity profile (solid line) for a solar metallicity K giant star (left) and a metal poor one (right). The dotted line is the intensity profile computed with the 1D model having identical stellar parameters, input
data, and chemical compositions as the 3D simulation. Parametric full (dash-dotted line) and partial
      limb-darkening (triple dot-dashed line) are also shown. \emph{Bottom panels:} Visibility curves computed with Hankel transform with a fixed angular diameter of 2 mas for the above intensity profiles of the 3D simulation (solid line), 1D
       (dotted line), and fully limb darkening (dash-dotted line). A logarithm scale is used on y-axis. Synthetic visibilities in these plots are not realistic near the nulls but are intended only for model-to-model comparison.  Images from Chiavassa {\em et al.\/} (\cite{2010A&A...524A..93C}).
          }
       \label{ld_fits}
  \end{figure*}

Stellar parameters determined with interferometric techniques often
consist into a simple uniform disk model-fitting that, although
unphysical, has the advantage of immediately telling if a star is
resolved, after which limb-darkening corrections are
applied. Chiavassa {\em et al.\/} (\cite{2010A&A...524A..93C})
reported the effect on stellar parameter determination of 3D intensity
distribution with respect to 1D models. This has been carried out
using synthetic intensity profiles (Fig.~\ref{ld_fits}, top) from 3D
simulations with different stellar parameters and metellicities that
have been compared to 1D models with exactly the same parameters.

The visibility can be derived using 
$V_{\lambda}\left(B,\Theta\right)$ from the intensity profile 
$I\left(\lambda,\mu\right)$ using the Hankel integral:
\begin{equation}\label{hankel}
V_{\lambda}\left(B,\Theta\right)=\frac{1}{A}\int^1_0I\left(\lambda,\mu\right)J_0\left(\frac{\pi B\Theta}{\lambda}\sqrt{1-\mu^2}\right)\mu d\mu 
\end{equation}
where $\lambda$ is the wavelength in meters, $B$ is the baseline in
meters, $\Theta$ is an arbitrary angular diameter in radians (2 mas
here), $J_0$ the zeroth order of the Bessel function,
$\mu=cos(\theta)$ (with $\theta$ the angle between the line of sight
and the radial direction), and A the normalization factor:

\begin{equation}
A=\int^1_0I\left(\lambda,\mu\right)\mu d\mu
\end{equation}

For typical red giant stars having $4600 < T_{\rm{eff}} < 5100$~K, the
differences in angular diameters vary from about $-3.5\%$ to about
$1\%$ in the visible, and are roughly in the range between $-0.5$ and
$-1.5\%$ in the infrared (Table~\ref{table2}), the corresponding
change in effective temperature being $\Delta T_{\rm{eff}} /
T_{\rm{eff}} = 1 - \sqrt{ \Theta_{\rm{3D}}/ \Theta_{\rm{1D}}}$.

While the impact of the corrections in Table~\ref{table2} is usually
not dramatic, they are not negligible to properly set the zero point
of the effective temperature scale derived by mean of this fundamental
method. In particular in the visible, the (partially) resolved (very)
metal-poor stars, it is important to take those corrections into
account for a correct derivation of their diameters. A second point of
concern is the reliability of the existing catalogs of calibrator
stars for interferometry that are based on red giants: the formal high
accuracy (a fraction of 1$\%$) based on 1D models on diameter
determination (M{\'e}rand {\em et al.\/} \cite{2010A&A...517A..64M})
may be impacted by the correction reported in Table~\ref{table2}.

Interferometry is advantageous in that it provides the ability to
directly measure stellar angular diameters and effective
temperatures. However, there is a non-negligible and still partly
unexplored model dependence (Casagrande {\em et
al.\/},\cite{2014MNRAS.439.2060C}). 

More details about the link between stellar parameters, interferometry
and asteroseismology are reported in Creevey's contribution to this
book.

\begin{table}
\begin{minipage}[t]{\columnwidth}
\caption{Ratio between the limb-darkened diameters recovered using 
 1D  models ($\Theta_{\rm{1D}}$) or 3D simulations
 ($\Theta_{\rm{3D}}$) and the corresponding 
change in effective temperature $\Delta T_{\rm{eff}}$ for the RHD simulations. Table from Chiavassa {\em et al.\/} (\cite{2010A&A...524A..93C}).}
\label{table2}
\centering
\renewcommand{\footnoterule}{}  
\begin{tabular}{c|cc|c|c}
\hline \hline
$\lambda$ [$\mu$m] &  [Fe/H]   &  ${\rm log}~g$  & $\Theta_{\rm{3D}}/\Theta_{\rm{1D}}$ & $\Delta T_{\rm{eff}}$ [K]\\
\hline
0.5 \footnote{central wavelength of the corresponding visible filter used in Chiavassa {\em et al.\/} (\cite{2010A&A...524A..93C})} & 0.0 &  2.2  & 1.003 & $-7$\\
2.14 \footnote{central wavelength of the corresponding FLUOR (K band) filter used in Chiavassa {\em et al.\/} (\cite{2010A&A...524A..93C})}& 0.0 & 2.2 & 0.996  & 9\\
\hline
0.5 & $-$1.0  &  2.2   & 0.991 & 21 \\
2.14& $-$1.0 & 2.2  &  0.996 & 9\\
\hline
0.5 & $-$2.0  &  2.2  & 0.982 & 46\\
2.14& $-$2.0 & 2.2  & 0.990 & 25 \\
\hline
0.5 & $-$3.0  &  2.2  & 1.011 & $-28$\\
2.14& $-$3.0 & 2.2  & 0.989  & 28\\
\hline
0.5 & $-$3.0  &  1.6  & 0.965 & 82 \\
2.14& $-$3.0 & 1.6 & 0.984 & 37 \\
\hline
\end{tabular}
\end{minipage}
\end{table} 

\section[Chromosphere]{Additional notes on chromospheric stellar activity with interferometry}

The chromosphere (in the classical, stratified, and highly
oversimplified view) is an intermediate region in the atmosphere of a
star, lying above the photosphere and below the corona. Chromospheric
activity, which encompasses diverse phenomena that produce emission in
excess of that expected from a radiative equilibrium atmosphere, is
tightly linked to changes in the stellar magnetic field, whether
periodic or irregular, and is therefore tied to the structure of the
subsurface convection zone, the starÕs rotation, and the
regeneration of the magnetic field via a self-sustaining dynamo
(Hall,\cite{2008LRSP....5....2H}).

Activity signature may manifest in changes in the radiative transfer
of lines formed very high in the atmosphere of a star, typically for
spectral lines sensitive to the increase in temperature of the
atmosphere. The hydrogen Balmer lines and the doublet H$\&$K of the Ca
II are used to estimate the strength of the chromospheric activity. Mg
II lines in the ultraviolet can also help in this direction.

The advent of a new instruments (VEGA at CHARA), working in near IR
and visible, opened the possibility to probe and characterize the
chromosphere using interferometry: for this purpose H$\alpha$ and Ca
II triplet lines have been observed using VEGA to extract the
signature of stellar activity of a late-type K giant star (Berio {\em
et al.\/}, \cite{2011A&A...535A..59B}). The authors managed to
determine the physical extents of the chromosphere of the star
measuring the ratio of the radii of the photosphere with respect to
the chromosphere using the interferometric measurements in the
H$\alpha$ and the Ca II infrared triplet line cores. These
interferometric measurements are unique and crucial to constrain
chromospheric models.

   \bibliographystyle{astro}

\end{document}